\documentclass{myelsart}

\usepackage{amssymb}

\usepackage{booktabs}
\usepackage{hepnicenames,graphicx,colordvi,gnuplotcolors}
\usepackage[dvips]{color}
\usepackage{rotating}

\newcounter{bla}

\input paperdef

\begin{document}

\begin{frontmatter}

\begin{flushright}
DCPT/08/172\\
IPPP/08/86\\
BONN-TH-2008-17
\end{flushright}

\title{{\tt {HiggsBounds}}: 
Confronting Arbitrary Higgs Sectors with Exclusion Bounds\\
from LEP and the Tevatron
}

\author[DESY]{P.~Bechtle},
\author[DURHAM]{O.~Brein},
\author[SANTANDER]{S.~Heinemeyer},
\author[DURHAM]{G.~Weiglein}\\
and
\author[BONN]{K.E.~Williams}.

\address[DESY]{DESY, Notkestrasse 85, 22607 Hamburg, Germany}
\address[DURHAM]{Institute for Particle Physics Phenomenology,
Durham University,\\ 
Durham, DH1 3LE, United Kingdom}
\address[SANTANDER]{Instituto de Fisica de Cantabria (CSIC-UC), 
Santander, Spain}
\address[BONN]{Bethe Center for Theoretical Physics, 
Physikalisches Institut der
Universit\"at Bonn,
Nussallee 12, 53115 Bonn, Germany}

\begin{abstract}
{\tt HiggsBounds} is a 
computer code
that
tests theoretical predictions
of models with arbitrary Higgs sectors against the exclusion bounds
obtained from the Higgs searches at LEP and the Tevatron. 
The included 
experimental information comprises exclusion bounds at 95\% C.L.\ on 
topological cross sections. 
In order to determine which search topology has the highest exclusion power,
the program also includes, for each topology, information from the 
experiments on the expected exclusion bound, which would have been
observed in case of a pure background distribution.
Using the predictions of the desired model provided by the user as input,
{\tt HiggsBounds} 
determines the most sensitive channel and tests whether the 
considered parameter point is excluded at the 95\% C.L.
{\tt HiggsBounds} is 
available as a Fortran 77 and Fortran 90 code.
The code can be invoked as a command line version, a subroutine version
and an online version.
Examples of exclusion bounds obtained with 
{\tt HiggsBounds} are discussed for the Standard Model, for 
a model with a fourth generation of
quarks and leptons and for the Minimal Supersymmetric Standard Model
with and without $\cp$-violation. The experimental information on the
exclusion bounds currently implemented in {\tt HiggsBounds} will be
updated as new results from the Higgs searches become available.
\end{abstract}

\end{frontmatter}

{\bf PROGRAM SUMMARY}

\begin{small}
\noindent
{\em Manuscript Title:} {\tt HiggsBounds}:
Confronting Arbitrary Higgs Sectors with Exclusion Bounds
from LEP and the Tevatron 					\\
{\em Authors:} P.~Bechtle, O.~Brein, S.~Heinemeyer, G.~Weiglein, K.E.~Williams\\
{\em Program Title:} {\tt HiggsBounds}                                         \\
{\em Journal Reference:}                                      \\
{\em Catalogue identifier:}                                   \\
{\em Licensing provisions:}                                   \\
{\em Program requirements:}
{\tt HiggsBounds} can be built with any compatible Fortran 77 or Fortran 90
compiler.
The program has been tested on x86 CPUs running under Linux (Ubuntu 8.04)
and with the following compilers: 
The Portland Group Inc. Fortran compilers (pgf77, pgf90),
the GNU project Fortran compilers (g77, gfortran).
\\
{\em Programming language:} Fortran 77, Fortran 90 (two code versions are offered) \\
{\em RAM:} minimum of about 6000 kbytes (dependent on the code version) 
{\em Classification:}                                         \\
{\em External routines/libraries:}                                      
{\tt HiggsBounds} requires no external routines/libraries.
Some sample programs in the distribution require
the programs FeynHiggs 2.6.x or CPsuperH2 to be installed.
\\
{\em No. of lines in distributed program, including test data, etc.:} 61645
\\
{\em No. of bytes in distributed program, including test data, etc.:} 13721912
\\
{\em Distribution format:} tar.gz
\\
{\em Nature of problem:}
Determine whether a parameter point of a given model
is excluded or allowed by LEP and Tevatron Higgs-boson search results.
\\
{\em Solution method:}
The most sensitive channel from LEP and Tevatron searches
is determined and subsequently applied to test this parameter point. 
The test requires as input model predictions for the Higgs-boson masses, 
branching ratios and ratios of production cross sections with respect
to reference values.
\\
{\em Restrictions:}
In the current version, results from decay-mode independent Higgs
searches and 
results of searches for charged Higgs bosons are not taken into account.
   \\
{\em Running time:}
about 0.01 seconds (or less) for one parameter point 
using one processor of an Intel Core 2 Quad Q6600 CPU at 2.40GHz
for sample model scenarios with three Higgs bosons.
It depends on the complexity of the Higgs sector 
(e.g. the number of Higgs bosons and the number of open decay channels)
and on the code version.


\end{small}

\newpage



\hspace{1pc}
{\bf LONG WRITE-UP}

\section{Introduction}
\label{sec:intro}

A major goal of the 
particle physics programme at the high energy frontier,
currently being pursued at the Fermilab Tevatron collider
and soon to be taken up by the CERN Large Hadron Collider (LHC),
is to unravel the nature of electroweak symmetry breaking.
While the existence of the massive electroweak gauge bosons ($W^\pm,Z$),
together with the successful description of their behaviour by
non-abelian gauge theory, 
requires some form of electroweak symmetry breaking to be present in nature, 
the underlying dynamics is not known yet.
An appealing theoretical suggestion for such dynamics is the Higgs mechanism
\cite{higgs-mechanism}, which 
implies the existence of one or more 
Higgs bosons (depending on the specific model considered).
Therefore, the search for Higgs bosons is a major cornerstone
in the physics programmes of past, present and future 
high energy colliders.

Many theoretical models employing the Higgs mechanism in
order to account for electroweak symmetry breaking
have been studied in the literature, of which 
the most popular ones are the Standard Model (SM)~\cite{sm}  
and the Minimal Supersymmetric Standard Model (MSSM)~\cite{mssm}.
Within the SM, the Higgs boson is the last undiscovered particle, whereas the
MSSM has a richer Higgs sector, containing three neutral and two charged
Higgs bosons. 
Among alternative theoretical models beyond the SM which lead to
the prediction of Higgs bosons, the most prominent are  
the Two Higgs Doublet Model (THDM)~\cite{thdm}, 
non-minimal supersymmetric extensions of the SM 
(e.g. extensions of the MSSM by an extra singlet
superfield \cite{NMSSM-etc}),
little Higgs models~\cite{lhm} and 
models with more than three spatial dimensions~\cite{edm}. 
Furthermore, several models have been proposed 
which mainly extend the Higgs sector of the SM,
e.g. extensions of the SM by electroweak singlet
scalar fields \cite{higgs-singlet-models}, 
or the Private Higgs Model \cite{private-higgs}
which introduces one Higgs doublet for each fermion generation,
or models with Higgs-dependent Yukawa couplings \cite{higgs-dep-Yuakawa}.

LEP has searched for the SM Higgs boson~\cite{LEP-SM-Higgs-analysis},
the MSSM Higgs bosons~\cite{LEP-MSSM-Higgs-analysis}, and
other, more ``exotic'', manifestations of the Higgs 
mechanism~\cite{LEPother}. After the
termination of LEP, the search has been taken up again by the Tevatron.
So far, no signals of Higgs bosons have been found, and LEP and 
Tevatron turned the non-observation of Higgs signals into
cross section constraints.
The constraints 
are provided by experiments in the form
of limits on cross sections of 
individual signal topologies (such as $e^+e^-\to h_iZ\to b\bar{b}Z$
or $p\bar p\to h_i Z\to b\bar b l^+l^-$) or in the form of combined limits for a
specific model, such as the SM. In the latter case,
the individual topological cross sections have been combined using the
proportions of the individual 
contributions
as predicted by the model. 

An important part of the phenomenological exploration of 
physics models beyond the SM is the 
confrontation of such models with existing experimental
constraints. In particular, checking the predictions of the model's
Higgs sector with bounds from experimental searches for a Higgs boson
is vital. 
In order to test whether the parameter space of a certain
model is compatible with the limits from the Higgs searches at LEP and
the Tevatron, it is, in general, not possible to employ the experimental
results that have been obtained for the Higgs search in the SM (or any other
specific model). This is because, in a general model of new physics,
individual Higgs signal topologies will contribute in
different proportions than, for instance, in the SM. Thus, for models for which 
no explicit experimental analysis has been carried out, or, like in the
case of the MSSM, for model parameters that differ from the benchmark values
chosen in the experimental analyses, one needs to resort to the limits
on the individual topological cross sections provided by the experimental
collaborations. These limits exist for various search channels from LEP
and the Tevatron (the latter
ones are frequently updated as new data becomes available). 
Comparing the predictions of a particular
model with the existing experimental bounds on the various 
search topologies can be
quite a tedious task as it involves the implementation of experimental
results that are distributed over many different publications and 
combining these results
requires a procedure to ensure the correct statistical interpretation of
the exclusion bounds obtained on the parameter space of the model.

We present here the program {\tt HiggsBounds}, which is a tool
designed to facilitate the above task so that wide classes of models can
easily be checked against the state-of-the-art results from Higgs
searches. This should be useful for applications in Higgs phenomenology
and model building. 
{\tt HiggsBounds} takes theoretical Higgs sector predictions,
e.g.\ for a particular parameter scenario of a model beyond the SM,
as input and determines which Higgs 
search analysis
has the highest exclusion power according to a list of expected
exclusion limits from LEP and the Tevatron (an expected exclusion limit 
corresponds to the bound that one would obtain in the hypothetical
case of an observed distribution that agrees precisely with the
background expectation).
In order to ensure the correct statistical
interpretation of the obtained exclusion bound as a 95\% C.L., the
comparison of the model with the experimental limits
has to be restricted to the single channel that possesses the 
highest statistical sensitivity for setting an exclusion limit.
For this channel, the program then compares the theoretical prediction 
for the Higgs production cross section 
times decay branching ratio
with the actual experimental limit and determines whether or not
the considered parameter point of the model is excluded at 95\% C.L.
So far, no bounds on charged Higgs bosons are taken into
account. This will be included in a later version.
{\tt HiggsBounds} is model-independent, in that it can be used for models 
with arbitrary Higgs sectors.
The input needed is the number of neutral Higgs bosons in the
model
and  masses, decay branching ratios and ratios of production cross sections
versus corresponding reference (usually SM) values for all neutral Higgs
bosons in the model. 
The code has both a 
Fortran 77 and Fortran 90 version. It can be operated in a command line
mode that can process
input files in a variety of formats, as a subroutine suitable for
inclusion in user applications, and as an online version, available 
at the URL:
{\tt http://www.ippp.dur.ac.uk/HiggsBounds}.
{\tt HiggsBounds} includes 
sample programs which demonstrate its usage.
Sample programs which demonstrate how {\tt HiggsBounds}
can be used in conjunction with the
widely used programs 
{\tt FeynHiggs}~\cite{feynhiggs,mhiggslong,mhiggsAEC,mhcMSSMlong} and 
{\tt CPsuperH}~\cite{cpsh} for
Higgs-sector predictions in the MSSM are provided by default.

The rest of the paper is organised as follows.
In Section \ref{sec:implementation} we describe our 
general approach for the implementation
of experimental Higgs constraints followed by specific details
on the implementation of LEP and Tevatron results.
In Section \ref{sec:usage}, the general operating instructions 
for {\tt HiggsBounds}
are specified and accompanied by small sample programs,
while in Section 
\ref{sec:examples} some examples of use for {\tt HiggsBounds}
are described. 
Section \ref{sec:summary} contains a summary and outlook
and is followed by an appendix which contains details on the definition 
and statistical interpretation of the confidence limits.

\section{Implementation of experimental Higgs constraints}
\label{sec:implementation}

\subsection{General approach}
\label{subsec:general approach}

Tevatron and LEP turn(ed) the non-observation of Higgs signals into
95\% C.L.\ upper limits on cross sections for individual signal topologies.
These limits are given as functions of the anticipated mass of the 
Higgs boson. By a certain Higgs signal topology $X$, we mean a specific 
combination of a single (double) Higgs boson production process, 
$P(h)$ ($P(h_1,h_2)$), and decay final state(s) $F$ of 
the Higgs boson(s) $h$ ($h_1,h_2$).
In the limit of a narrow-width Higgs boson
the cross section $\sigma(X)$ of the signal topology $X$ factorises
into the on-shell Higgs production cross sections $\sigma(P)$
times the appropriate Higgs decay branching ratio(s) $\BR(h\to F)$
($\BR(h_1,h_2\to F)$).

The Tevatron and LEP cross section limits are understood 
to be applicable to models which do not 
change the signature of the background processes considerably\footnote{ 
This is not such a strong extra restriction on new physics models
as it may seem, because viable models are required to 
agree with present experimental results which are well described by
the SM, such as the LEP and SLD EW precision data~\cite{:2005ema} and the 
precision measurements of gauge boson production at LEP 
and the Tevatron \cite{Alcaraz:2006mx}. These analyses are much more
sensitive to changes in the  
background to Higgs searches than the Higgs searches themselves.
Therefore, the models with new physics in the Higgs sector 
which have been considered in
the literature usually do not show strong
deviations from the SM in the background processes.}
which do not significantly change the kinematical distributions of the  
signal topology $X$ (e.g.\ $\eta$, $p_T$ distributions of the final state 
particles) from what has been assumed in the
corresponding analysis.  This is an unavoidable limitation of the standard
approach adopted in the literature of expressing search results in terms  
of limits on total cross sections times branching ratios for certain
signal topologies.
Detection efficiencies, for instance the b-jet
tagging efficiency, usually depend on the kinematical configuration
of the signal event (particularly the $\eta$ and $p_T$ of the final
state particles).
Therefore, a dedicated experimental analysis for a model giving rise
to kinematical distributions that strongly differ from the ones assumed in
the analysis on which the published cross section limit is based would 
lead to a somewhat modified cross section limit.
For instance, applying the cross section limit of a Tevatron analysis 
which relies on b-jet tagging to a model where the distribution of the 
signal b-jets differs strongly from the one assumed in the Tevatron
analysis would lead to a slightly too weak/strong limit if a higher/lower
fraction of the signal b-jets than assumed in the analysis is radiated 
into the central part of the detector where the
b-jet tagging efficiency is higher than in regions closer in angle
to the beam axis.
{\tt HiggsBounds} provides three options for passing model predictions to
the program: ({\tt hadr}, {\tt part} and {\tt effC}). The extent to which
the kinematical distributions may vary from those assumed in the analyses
depends on which of these input options is chosen.
We will discuss this issue in further detail in
the sections describing the input options.

The Tevatron and LEP cross section limits 
are usually given for a narrow-width Higgs boson.
From the narrow-width limits, some experimental analyses 
extract limits for Higgs bosons with a substantial decay width
by representing the signal of a ``large-width'' Higgs boson
as a weighted sum of mock narrow-width Higgs signals. 
Examples of this procedure can be found in
\cite{CDF-3b-analysis1,CDF-3b-analysis2,D0-0805-3556,D0-0805-2491}.
In particular, \cite{D0-0805-2491} shows 
that the effect of a non-vanishing width on the extracted limit
can be parametrised by a function of the Higgs mass $m_h$ 
and the ratio $\Gamma_\TOT(h)/m_h$.
If model-independent information of this kind also became available
(for a fine grid of points)
for other analyses, this
would greatly facilitate the task to implement width-dependent 
Higgs exclusion limits.

At present, we constrain the implementation of {\tt HiggsBounds}
to the case of narrow-width Higgs boson exclusion limits. 
Neglecting the width of a Higgs boson usually leads to 
tighter bounds on the topological cross sections
than those obtained through taking the width into account.
The user should be aware of this fact when considering exclusion bounds 
on parameter regions that give rise to very large Higgs-boson widths. 
In a future 
version of {\tt HiggsBounds} we plan to incorporate a proper width
treatment for cases where width-dependent
results are provided by the experimental collaborations. 
As mentioned above, {\tt HiggsBounds} is not meant to be applied to
models for which the background rates to the considered signal processes
differ drastically to the SM case.

For some specific models (the SM and some
benchmark scenarios of the MSSM) the results of 
several Higgs signal topologies have been 
combined 
by the experimental collaborations
according to the relative contributions
to the total cross section.
In the combination, a detailed knowledge of the overlap between the
individual experimental searches is used, which is not available using
the limits on topological cross sections only. 
Hence, a combined limit, for instance for the SM,
is more sensitive than a combination of individual topological
limits. 
The use of detailed SM dependent information during
the calculation of the combined limit prohibits its application to
those Higgs bosons of new physics models 
which are not predicted to have all relevant couplings
proportional to the SM values. 
However, such models can still contain one (or some)
Higgs boson(s) which are SM-like.
For those Higgs bosons, the SM combined 
limit can be applied.

The basic flow of {\tt HiggsBounds} is as follows.
\begin{enumerate}
\item 
The user provides (in a form to be specified below)
for all neutral Higgs bosons $h_i \; (i=1,\ldots,n_\HIGGS)$
of the considered model the predictions for:
Higgs masses, 
total decay widths, 
the relevant branching ratios for the decay of Higgs bosons $h_i$
into known ordinary particles (OP, i.e.\ SM fermions and gauge bosons)
and 
into pairs of other Higgs bosons $h_j$,
and
the relevant cross sections for Higgs production processes $P$ 
normalised to reference (usually SM) values:
\begin{align}
\label{basic input}
& m_{h_i} \,, 
\Gamma_\TOT(h_i)\,,  
\BR(h_i \to \text{OP}) \,, 
 \BR(h_i \to h_j h_j)\,,\\
\nonumber
& \left.\frac{\sigma_\MOD(P(h_i))}{\sigma_\REF(P(H),m_H)}
	\right|_{m_H=m_{h_i}}\,,
  \left.\frac{\sigma_\MOD(P(h_i,h_j))}{\sigma_\REF(P(H,H'),m_{H},m_{H'})}
	\right|_{m_{H,H'}=m_{h_{i,j}}}\,.
\end{align}
The precise definitions of the reference cross sections are 
given in the LEP- and Tevatron-specific subsections below.
The user can choose to specify all the above quantities directly, or, 
in the simplest form, one just needs to provide $m_{h_i},
\Gamma_\TOT(h_i), \BR(h_i \to h_j h_j)$ and
effective couplings of the Higgs bosons to ordinary particles.
The remaining branching ratios and cross section ratios listed above 
can then be calculated from this input in the effective coupling 
approximation.
For the preparation of normalised input, the user may apply
the SM predictions for Higgs boson production cross sections
and decay branching ratios which {\tt HiggsBounds} provides 
as external routines (see Table \ref{SMfunctions} 
in Section \ref{subsec:Library of subroutines}).

\item 
With the above model input,
the program calculates the 
model predictions for the 
cross section (normalised to a reference value in some cases) 
for each Higgs signal topology 
$Q_\MOD(X)$. 
Depending on whether the exclusion result
for a particular search topology has been
given by the experimental collaborations
as a relative or absolute limit, the 
program evaluates
\begin{align*}
Q_\MOD &= \frac{[\sigma\times\BR]_\MOD}{[\sigma\times\BR]_\REF}
\;\;\text{or}\;\; [\sigma\times\BR]_\MOD
\,,
\end{align*}
where $\sigma$ and $\BR$ denote the production process 
and decay branching ratio of the Higgs boson, respectively.

\item 
From the experimental results, we read off
the values $Q_\OBS(X)$ and $Q_\EXPEC(X)$
corresponding to the observed and expected 
95\% C.L. exclusion bound on the quantity $Q(X)$, respectively.
To that end, we have implemented 
many of the available observed and expected
exclusion limits (preliminary and final results)
from 
LEP and the Tevatron 
as data tables which are read in by the program during 
start-up.
In order to 
provide values for $Q_\OBS(X)$ and $Q_\EXPEC(X)$ for 
continuous Higgs mass values, the program interpolates $Q$-values
linearly between neighbouring Higgs mass points given by the 
experimental collaborations. 
Details on the definition of 95\% C.L. limits can be found in
Appendix \ref{S95-LEP}.

\item 
The program determines the search topology $X_0$ 
with the highest ratio of $Q_\MOD$ to expected $Q_\EXPEC$,
i.e. out of all topologies $X$, it finds the channel $X_0$ where 
\begin{align*}
\frac{Q_\MOD(X)}{Q_\EXPEC(X)}
\end{align*}
is maximal.
This topology has the highest statistical sensitivity 
for excluding the model prediction.

\item
We then look at the ratio of $Q_\MOD$ to the 
observed $Q_\OBS$ value for this channel $X_0$ and, 
if
\begin{equation}
\frac{Q_\MOD(X_0)}{Q_\OBS(X_0)} > 1
\,,
\label{eq:modvsobs}
\end{equation} 
this particular parameter point is excluded 
by the corresponding
experimental analysis for the search channel $X_0$
at 95 \% C.L.
In order to ensure the correct statistical
interpretation of the obtained exclusion bound as a 95\% C.L., 
it is crucial to perform the test of \refeq{eq:modvsobs} only for
exactly one search channel (namely the one with the highest statistical
sensitivity), and not for all
available search channels. In the latter case the
derived constraint would in general not correspond to a constraint at 
95~\% C.L.,
due to the maximally 5~\% probability of each individual comparison
of $Q_\MOD$ and $Q_\OBS$ to yield a false exclusion.
\end{enumerate}

As additional information, the web version of the program {\tt HiggsBounds} also informs the
user which topologies had the second and third highest statistical
sensitivity and whether or not the ratio of $Q_\MOD$ to 
observed 95\% C.L. limit $Q_\OBS$ was greater than one for those topologies. 
However, this
information does not contribute to the decision about whether or not
this parameter point is excluded.

\bigskip

There are a few more general implementation details which 
should be mentioned here.

{\em Using SM combined limits, SM-likeness}\\
In order to use the best available exclusion limit, it makes sense to 
use the results of the SM combined analyses for any Higgs boson predicted 
to behave in a similar way to the SM Higgs boson 
(in terms of its production and decay).
A SM analysis, combining several signal topologies
assumes that the relative contribution to the event rate by each included
search topology is according to the proportions in the SM.
For a given Higgs mass $m_H$, the only free parameter left in such analyses
is one overall scale factor,
which multiplies all Higgs signal cross sections and 
which is varied in order to determine the 
value for which 95\% C.L. exclusion occurs.

Therefore, the results of a SM analysis, combining $M$ signal topologies, are 
exactly applicable to a Higgs boson $h$ of an alternative model
if for all of the $M$ search topologies $X_n$ the ratio of 
topological cross sections (model versus SM) is independent of $X_n$, i.e.
\begin{align}
\label{SM-likeness1}
\frac{\sigma_\MOD(X_n)}{\sigma_\SM(X_n)} & = \text{const.} =: 
\alpha \,, & n=1,\ldots,M 
\,,
\end{align}
with a proportionality constant $\alpha$, or equivalently
\begin{align}
\label{SM-likeness1b}
\frac{\sigma_\MOD(X_n)}{\sigma_\MOD(X_k)} 
	& = \frac{\sigma_\SM(X_n)}{\sigma_\SM(X_k)} 
	& \text{for all}\; n,k \in \{1,\ldots,M\}
\,.
\end{align}

For a model which passes the conditions (\ref{SM-likeness1}) or
(\ref{SM-likeness1b}) to sufficient accuracy
for a set of $M$ Higgs search topologies,
the cross section prediction
for a combination of those topologies,
normalised to the SM reference value, fulfils
\begin{align}
\label{def-Q-combined}
Q_\MOD(\{X_1,\ldots X_M\}) &:= 
	\frac{\sum_{n=1}^{M}\sigma_\MOD(X_n)}
	     {\sum_{n=1}^{M}\sigma_\SM(X_n)} = \alpha
\,.
\end{align}

Here and in the following, it is always understood that the mass of 
the SM Higgs boson $m_H$ is set to $m_h$.
In the narrow-width approximation, which we 
assume to hold throughout (see the discussion above),
each cross section $\sigma(X_n)$ 
of the Higgs signal topologies $X_n$
factorises into a production cross section $\sigma(P_i(h))$
and decay branching ratio $\BR(h\to~F_k)$,
where $i$ and $k$ are functions of $n$, and $n=1,\ldots,M$.
Note that the index $n$ labels distinct signal {\em topologies},
i.e.\ specific combinations of one Higgs production process with 
one decay final state.
SM-likeness is fulfilled if
\begin{align}
\label{SM-likeness2}
\frac{\sigma(P_i(h))_\MOD}{\sigma(P_i(H))_\SM}
\frac{ \BR_\MOD(h \to F_k)}{\BR_\SM(H \to F_k)} 
	& = \text{const.} 
\end{align}
holds to sufficient accuracy for all $M$ topologies.
If this is the case for a particular Higgs boson of a certain model,
a SM analysis combining precisely those $M$ channels also 
applies to that Higgs boson.

In order to decide whether or not a Higgs boson of a certain model is 
sufficiently SM-like such that the corresponding SM combined analysis
can be applied, we use in 
{\tt HiggsBounds} a pragmatic definition of SM-likeness.
For each of the $N_{\text{CS}}$ distinct production cross sections $\sigma_\MOD(P_i(h))$
and the $N_{\text{BR}}$ distinct decay branching ratios $\BR(h\to F_k)$
which appear in the list of $M$ signal topologies,
we determine the normalised mean value and the deviation from the mean:
\begin{align}
\bar s &= \frac{1}{N_{\text{CS}}} \sum_{i=1}^{N_\text{CS}} s_i
\,, & 
\bar b &= \frac{1}{N_{\text{BR}}} \sum_{k=1}^{N_\text{BR}} b_k
\,, \\
\delta s_i & = s_i - \bar s
\,, &
\delta b_k & = b_k - \bar b
\,,\\
\text{with}\;
s_i & = \frac{\sigma_\MOD(P_i(h))}{\sigma_\SM(P_i(H))}
\,, &
b_k & = \frac{\BR_\MOD(h\to F_k)}{\BR_\SM(H\to F_k)}
\,.
\end{align}
The approximate SM-likeness of the given model
is then ensured if the deviations from a constant expression
according to Eq.~(\ref{SM-likeness2}) are small.
For a given combination $(i,k)$ of a production and decay process,
the left hand side of  Eq.~(\ref{SM-likeness2}) can be
expressed as:
\begin{align}
\frac{\sigma_\MOD(P_i(h))}{\sigma_\SM(P_i(H))} 
\frac{\BR_\MOD(h\to F_k)}{\BR_\SM(H\to F_k)} = s_i b_k
	& = \bar s\, \bar b \left(1 + \frac{\delta s_i}{\bar s} 
	+ \frac{\delta b_k}{\bar b} 
	+ \frac{\delta s_i \delta b_k}{\bar s\, \bar b} 
	\right)
\,.
\end{align}
In the program, we consider a corresponding SM analysis applicable
if the maximum relative deviation from the 
mean value $\bar s\, \bar b$ satisfies
\begin{align}
\max_{i,k} \left(\frac{\delta s_i}{\bar s}
        + \frac{\delta b_k}{\bar b}
        + \frac{\delta s_i \delta b_k}{\bar s\, \bar b} \right) < \epsilon
\,, 
\end{align} 
where we have chosen $\epsilon = 2\%$, i.e.\ 
the predictions for the different topological cross sections, normalised
to the SM values, 
are required to be equal to each other with
at least 2\% accuracy. This restriction is quite stringent and may cause a
SM combined analysis to be not applied to some models where it may
actually be justifiable. 
The user can remedy this situation by changing the
setting of the corresponding variable {\tt eps} in the 
internal subroutine {\tt check\_SM\_likeness} (Fortran 77)
or {\tt SMlikeness} (Fortran 90) in the code.

So far, {\tt HiggsBounds} uses analyses with SM combinations of 
topologies only from the Tevatron experiments.
This is the most important case, as CDF and D\O\ have not done 
combinations of their results for individual search topologies so far, 
while the LEP results we are using represent the combination of results 
from all four LEP experiments for each individual channel.

A more elaborate criterion for determining whether SM
combined analyses are applicable to Higgs bosons of models beyond the SM can
be sketched as follows. The variation of the ratios of signal topology cross
sections (model versus SM) due to non-exact SM-likeness of the model
prediction should stay within a fraction of the 1$\sigma$ band of the
sensitivity of the observed limit of the analysis with respect to
signal-like fluctuations. The fixed maximum deviation of $2\%$, described above,
lies well within the typical expected statistical fluctuation of the 
limits in the SM combined analyses.
We plan to implement such a more elaborate criterion in 
future updates of the program.

{\em Adding of signal channels}\\
In models with more than one Higgs boson it can happen that
some of the Higgs bosons, $h_i$, have masses which are quite 
close to each other, say
\begin{align}
|m_{h_i}-m_{h_j}| & \leq \Delta m_h
\end{align}
for certain $i$ and $j$.
This is the case for instance in the parameter 
space of the ($\cp$-conserving) 
MSSM for the heavy Higgs bosons $H^0$ and $A^0$ for 
$\MA \gg \MZ$.
If the masses are so close to each other that 
their signals appear within a certain mass window,
it may happen that the two signals 
are not experimentally distinguishable anymore,
or, at least, that it makes sense to combine their signal 
contribution into one signal to be compared with the data.

In the narrow-width approximation 
(which is used in {\tt HiggsBounds}, see the discussion above) 
it is assumed that the total widths of the Higgs bosons $\Gamma_\TOT(h_i)$
are all (much) smaller than their mutual mass differences,
\begin{align}
\label{separate-peaks}
\Gamma_\TOT(h_i) & \ll |m_{h_i}-m_{h_k}| & \text{for all $i$ and $k\not=i$ }
\,.
\end{align}
In this case the signal channels factorise into production cross section
and decay branching ratio, so that the addition of different signal
cross sections may formally always be carried out where it is necessary. 
In parameter regions where the narrow-width approximation does not hold,
however, one may worry about the possible effects of interference terms
arising from squaring the amplitude of the 
full process involving both production and decay. It should be noted
that even in cases where the total widths of two Higgs bosons are
significantly larger than their mass difference it can happen that
interference terms are small. An example for such a situation is the 
above-mentioned case of the MSSM with real soft-breaking parameters
for $\MA \gg \MZ$.
While for large $\tb$ the widths of the two Higgs bosons can be very
large, the different quantum numbers of the $\cp$-even and the $\cp$-odd
state nevertheless forbid an interference contribution.

In determining the cross sections $\sigma(X)$
for a given signal topology $X$, {\tt Higgs\-Bounds} 
does the following.
Suppose there are $n_\HIGGS$ neutral Higgs bosons $h_i$ in a given model
and let the numbering be according to a mass order, i.e.
\begin{align}
m_{h_1}\leq m_{h_2} \leq \cdots \leq m_{h_{n_\HIGGS}}
\,.
\end{align}
Then, {\tt HiggsBounds} works through a loop from $i=1$ to $n_\HIGGS$
and adds up all\footnote{If the cross section for a SM combination of 
signal topologies is calculated, production cross sections of 
those Higgs bosons are only included in the sum if they
also fulfil the SM-likeness criterion.} 
production cross sections of Higgs bosons $h_j$ $(j\geq i)$ 
which satisfy 
\begin{align}
m_{h_j} - m_{h_i} \leq \Delta m_h(X)
\,.
\end{align}
In the implementation, the $X$-dependence of $\Delta m_h$
is limited to two different values: one for LEP and one for 
Tevatron search topologies, represented by the variables 
{\tt delta\_Mh\_LEP} and {\tt delta\_Mh\_TEV}, respectively.
Adding of Higgs signal cross sections is mainly relevant for Tevatron Higgs 
search analyses, where the invariant mass distributions of 
the potential Higgs boson decay products usually use bin sizes of the 
order of 10 GeV. 
For the LEP analyses, the Higgs mass resolution is typically around 
2--3 GeV \cite{LEP-SM-Higgs-analysis}. 
For the LEP results the 
addition of Higgs signal cross sections is implemented 
in {\tt HiggsBounds} only for the
Higgsstrahlung process where the Higgs boson decays directly into SM
particles. 
As default values for {\tt delta\_Mh\_LEP} and {\tt delta\_Mh\_TEV} 
we have chosen (in units of GeV):
\begin{align*}
\text{\tt delta\_Mh\_TEV} & = 10
\,,\\
\text{\tt delta\_Mh\_LEP} & = 2
\,.
\end{align*}
The default setting of {\tt delta\_Mh\_TEV} for the Tevatron search 
channels has also been tested against known MSSM Higgs search results 
\cite{D0-0805-3556,D0-0805-2491,CDF-9071}
which we could reproduce successfully with {\tt HiggsBounds} 
within the limitations of the narrow-width approach.

In the calculation of normalised quantities, the reference cross section
and branching ratio are evaluated at the average mass of the 
Higgs bosons contributing to a signal topology.

{\em Provided SM normalisation}\\
In the evaluation of the model predictions $Q_\MOD(X)$ and 
$Q_\MOD(\{X_1,\ldots,X_M\})$ for single and multiple search
topologies, respectively, to be compared with the 
expected and observed limits $Q_\OBS$ and $Q_\EXPEC$, 
{\tt HiggsBounds} uses internally SM predictions
for Higgs production cross sections and 
decay branching ratios.
Out of the set of implemented Higgs search results,
the need for using such predictions internally arises only 
for Tevatron analyses.
The provided SM predictions for Higgs boson production cross sections
and decay branching ratios use
results of  HDECAY 3.303 \cite{hdecay} and 
the TEV4LHC Higgs Working Group \cite{TEV4LHCWG-Higgs-CS}
(see Table \ref{SMfunctions} below for references to the original publications).
In addition,
for the $bH$ production cross section we need several predictions
with different kinematic cuts on the $b$ jet which are not 
available from \cite{TEV4LHCWG-Higgs-CS}.
We used HJET 1.1 \cite{HJET} to calculate those, but checked that 
the uncut cross section closely resembles the result reported in
\cite{TEV4LHCWG-Higgs-CS}
(see Section \ref{subsec:Library of subroutines} for details).
If the model input is given in terms of ratios of partonic cross sections,
the calculation of hadronic Higgs production cross sections 
is facilitated by using internally SM cross section ratios 
which we have calculated
(see Section \ref{subsubs:Tevatron-part-IO-part} for details).
If the model input is given in terms of 
effective couplings, we make use of results from 
VBFNLO \cite{VBFatNLO} in order to facilitate the 
calculation of the 
Higgs production cross section via vector boson fusion
(see Section \ref{subsubs:Tevatron-part-IO-effC} for details).

The rationale behind the choice of SM normalisation 
is that virtually all Tevatron analyses
implemented in this program use these predictions when normalising 
their cross section limits to SM quantities. 
Thus, describing deviations from the SM of a new model by 
using the SM normalisation the experimental analyses have chosen 
allows for the most accurate interpretation of the limits.
By using SM predictions 
for Higgs production cross sections which deviate 
from the internally used ones by a certain percentage
(be it because of a different loop order, 
different numerical values of input parameters,
renormalisation scheme or choice
of parton distribution functions),
a deviation of the same relative size will be caused in 
the quantities $Q_\MOD$.
The user should bear this in mind when interpreting the output of 
{\tt HiggsBounds}.

\subsection{LEP limits}
\label{LEP-limits}

The main Higgs boson production processes at LEP have been
Higgsstrahlung and
double Higgs production:
\begin{align}
e^+ e^- & \to h_k Z \,, & e^+ e^- & \to h_k h_i
\,.
\end{align}
Both processes are assumed to be mediated via $s$-channel $Z$ boson 
exchange in leading order of perturbation theory.

Currently, {\tt HiggsBounds} uses the exclusion limits on
topological cross sections 
obtained from the combination of all LEP2 and some LEP1
Higgs boson search results from Ref.~\cite{LEP-MSSM-Higgs-analysis}
which considers the Higgs decay final states $b\bar b$ and $\tau^+ \tau^-$.
Apart from SM search topologies, 
this analysis also
considers double Higgs production topologies, and
almost all possibilities for the decay of an on-shell Higgs boson 
into two lighter ones\footnote{
	The only possibilities not analysed in
	\cite{LEP-MSSM-Higgs-analysis} are the ones where 
	more than two distinct Higgs bosons appear. For example, 
	this occurs if 
	a primarily produced Higgs boson $h_k$ 
	decays into two different Higgs bosons $h_j$ and $h_i$,
	or if two Higgs bosons $h_k h_j$ are primarily produced 
	and one of these,
	say $h_k$,  decays into two Higgs bosons $h_i$ ($i,j,k$ all
	different).
	We are not aware of any
	studies by the LEP experiments
	of those more general topologies.
	} 
and their subsequent decay into matter particles.
Because of the limitation in \cite{LEP-MSSM-Higgs-analysis}
to $b\bar b$ and $\tau^+ \tau^-$ final states,
Higgs bosons which decay predominantly invisibly 
cannot be well constrained by {\tt HiggsBounds} at present.
We will include the appropriate LEP results 
in future updates of the code.
The search topologies included in {\tt HiggsBounds} up to now are 
shown in Table~\ref{LEP-search-toplogies}.

\begin{table}
\begin{tabular}{l|l}
search topology $X$ & reference\\
\hline
$e^+e^-\to (h_k)Z\to (b \bar{b})Z$ &  
	\cite{LEP-MSSM-Higgs-analysis,commun-Read} table 14b\\
$e^+e^-\to (h_k)Z\to (\tau^+ \tau^-)Z$ & 
	\cite{LEP-MSSM-Higgs-analysis,commun-Read} table 14c\\
$e^+e^-\to (h_k\to h_i h_i)Z\to (b \bar{b} b \bar{b})Z$ &  
	\cite{LEP-MSSM-Higgs-analysis,commun-Bechtle} table 15\\
$e^+e^-\to (h_k\to h_i h_i)Z\to (\tau^+ \tau^- \tau^+ \tau^-)Z$ & 
	\cite{LEP-MSSM-Higgs-analysis,commun-Bechtle} table 16\\
$e^+e^-\to (h_k h_i)\to (b \bar{b} b \bar{b})$ & 
	\cite{LEP-MSSM-Higgs-analysis,commun-Bechtle} table 18\\
$e^+e^-\to (h_k h_i)\to (\tau^+ \tau^- \tau^+ \tau^-)$ &  
	\cite{LEP-MSSM-Higgs-analysis,commun-Bechtle} table 19\\
$e^+e^-\to (h_k\to h_i h_i)h_i\to (b \bar{b} b \bar{b})b \bar{b}$ & 
	\cite{LEP-MSSM-Higgs-analysis,commun-Bechtle} table 20\\
$e^+e^-\to (h_k\to h_i h_i)h_i\to (\tau^+ \tau^- \tau^+ \tau^-)\tau^+ \tau^-$ &  
	\cite{LEP-MSSM-Higgs-analysis,commun-Bechtle} table 21\\
$e^+e^-\to (h_k\to h_i h_i)Z\to (b \bar{b})(\tau^+ \tau^-)Z$ &  
	\cite{commun-Bechtle}\\
$e^+e^-\to (h_k\to b \bar{b})(h_i\to \tau^+ \tau^-)$ & 
	\cite{commun-Bechtle}\\
$e^+e^-\to (h_k\to \tau^+ \tau^-)(h_i\to b \bar{b})$ & 
	\cite{commun-Bechtle}
\end{tabular}  \\
\caption{\sl\label{LEP-search-toplogies}LEP search topologies used by 
{\tt HiggsBounds}. For each Higgs search topology which has been
studied in \cite{LEP-MSSM-Higgs-analysis}, 
we indicate the corresponding table number.
It is assumed in all tables that $m_{h_k} > m_{h_i}$.
In the program, we use more fine-grained results 
\cite{commun-Read,commun-Bechtle}
from the analysis \cite{LEP-MSSM-Higgs-analysis} 
than have been displayed as tables in 
\cite{LEP-MSSM-Higgs-analysis}.
We also use equally fine-grained results 
for the expected limits \cite{commun-Read,commun-Bechtle},
which have only been displayed in graphs in \cite{LEP-MSSM-Higgs-analysis},
and for the observed and expected limits of a few channels, which
were not 
published \cite{commun-Bechtle}.
\vspace*{5mm}}
\end{table}

The limits on topological cross sections in \cite{LEP-MSSM-Higgs-analysis}
have been given in the normalisation where all Higgs decay branching 
ratios are equal to 1. 
For given Higgs mass(es), limits on the production cross sections
of the individual search topologies $X$
have been derived for this normalisation
in the form of scaling factors $S_{95}(X)$, defined as
\begin{align} 
	S_{95}(X)&=\sigma_{\rm max}(X)/\sigma_{\rm ref}(X), 
\end{align}
where $\sigma_{\rm max}$ is the largest cross section compatible with the
data at 95 \% C.L. and $\sigma_{\rm ref}$ is a reference cross section for the
Higgs production process.
The quantities $S_{95}(X)$ of \cite{LEP-MSSM-Higgs-analysis}
correspond to the quantities $Q(X)$ mentioned in the description
of the program flow in Section~\ref{subsec:general approach}.

For Higgsstrahlung processes,
$e^+e^-\to h_k Z\to (\text{final state})Z$, 
the reference cross section $\sigma_{\text{ref}}$ is the SM cross section 
\begin{align}
\label{eeHZ-ref}
\sigma_{\text{ref}}(\text{Higgsstrahlung})& =\sigma(e^+e^-\to H Z)_{\rm SM}
\end{align}
with the SM Higgs mass $m_H$ chosen as $m_{h_k}$.

For double Higgs production processes, 
$e^+e^-\to H'H\to (\text{final state})$,
there is no direct SM reference process.
However, as in \cite{LEP-MSSM-Higgs-analysis},
we choose a reference cross section for a fictitious 
production process of two scalar particles ($H'$, $H$) with masses 
$m_{H'}=m_{h_k}$ and 
$m_H = m_{h_i}$ via a virtual
$Z$ exchange with a standardised coupling 
constant 
\begin{align}
\label{gref-HHZ}
g^\REF_{H'HZ} =\frac{e}{2\, \sw \cw}
\,,
\end{align}
where $e$ denotes
the electromagnetic coupling constant, and $\sw$ and $\cw$
the sine and cosine
of the electroweak mixing angle, respectively.
The cross section of this process in leading order is then completely 
determined by the Higgs masses and SM input and related
to the SM Higgsstrahlung cross section via a simple phase space factor:
\begin{align}
\label{eeHH-ref}
\sigma_\REF(\text{$H'\, H$ production})&
	=\bar{\lambda}\left(m_{H'},m_{H},s\right)
         \sigma^{\rm SM}_{HZ}(m_H) \,,\\
\nonumber
\bar{\lambda}\left(m_{H'},m_{H},s\right)&
	=\frac{\lambda^{3/2}_{H'H}(s)}
        {\lambda^{1/2}_{HZ}(s)\left(\lambda_{HZ}(s)+12
         \frac{m_Z^2}{s}\right)} \,,\\
\nonumber
\lambda_{ab}(s)&=\left[1-\frac{\left(m_a+m_b\right)^2}{s}\right]
                 \left[1-\frac{\left(m_a-m_b\right)^2}{s}\right] \,.
\end{align}
This reference cross section 
coincides with the MSSM or THDM tree-level
cross section 
for the process $e^+ e^- \to h^0 A^0$ if the Higgs mixing-angle dependent 
factor $\cos(\beta-\alpha)$ is divided out and $m_{A^0}$ and $m_{h^0}$ 
are chosen as $m_{H'}$ and $m_{H}$.

In order to apply the LEP Higgs constraints implemented in {\tt HiggsBounds} 
to a model, the user has to provide model information which 
allows  model predictions to be calculated
for all Higgs search topologies $X$ 
listed in Table \ref{LEP-search-toplogies},
\begin{align}
Q_\MOD(X) & = \frac{\sigma_\MOD(P(X))\BR_\MOD(F(X))}{\sigma_\REF(P(X))}
\,,
\end{align}
where the symbol $P(X)$ denotes 
the corresponding Higgs production process and $F(X)$
the corresponding Higgs decay chain,
e.g. $h_k\to b\bar b$ or $h_k\to h_i h_i\to b\bar b\tau^+\tau^-$.
The quantity $Q_\MOD(X)$ is compared with 
the experimental limits $Q_\EXPEC(X)$ and $Q_\OBS(X)$ according to the 
description in Section~\ref{subsec:general approach}. 
Currently, {\tt HiggsBounds} allows
model information to be provided in three ways.
\medskip

\subsubsection{LEP input ({HiggsBounds} Input Option {\tt part} and {\tt hadr}): 
cross section and branching ratios}
\label{subsubs:LEP-part-IO-part-hadr}
In order to be able to apply the implemented LEP limits to
the model under study,
the user is asked to provide model predictions for
the Higgs boson masses $m_{h_k}$
(in units of GeV), 
for the branching ratios
\begin{align}
\label{IO-full-LEP-BRs}
 \BR_\MOD(h_i\to b\bar b)\,, & & \BR_\MOD(h_i\to \tau^+ \tau^-) \,, 
	& & \BR_\MOD(h_k\to h_i h_i) \,,
\end{align}
and for the ratios of cross sections
\begin{align}
\label{IO-full-LEP-CSs}
& \frac{\sigma_\MOD(e^+e^-\to h_k Z)}
	{\sigma_\REF(e^+e^-\to H Z)}\,, 
& & \frac{\sigma_\MOD(e^+e^-\to h_k h_i)}
	{\sigma_\REF(e^+e^-\to H' H)}\,,
\end{align}
for $k,i \in \{1,\ldots,n_\HIGGS\}$.

With this information, {\tt HiggsBounds} obtains
model predictions $Q_\MOD(X)$ for all topologies $X$
specified in Table~\ref{LEP-search-toplogies}.
For example, the $Q_\MOD$ value for 
the channel $e^+e^- \to h_1 Z\to (b \bar{b})Z$ 
is given by (suppressing the initial state in the notation):
\begin{align*}
Q_\MOD\left((h_1) Z\to (b \bar{b})Z\right)&=
	\frac{\sigma_{\rm model}(h_1 Z)}{\sigma_\REF(H Z)} 
	\; \BR_\MOD(h_1\to b \bar{b})
\,,
\end{align*}
with $m_H = m_{h_1}$, and similarly
for the channel 
$e^+e^-\to (h_2\to h_1 h_1)Z\to (b \bar{b} b \bar{b})Z$ 
it is:
\begin{multline*}
Q_\MOD\left((h_2\to h_1 h_1)Z
	\to (b \bar{b} b \bar{b})Z\right)=\\
	\frac{\sigma_\MOD(h_2 Z)}{\sigma_\REF(H Z)}
	\BR_\MOD(h_2\to h_1 h_1)
	\left(\BR_\MOD(h_1\to b \bar{b})\right)^2
\,,
\end{multline*}
with $m_H = m_{h_2}$.

The {\tt HiggsBounds} Input Option {\tt hadr}
amounts to exactly the same LEP related input  as 
described above.
It differs for the Tevatron related input 
(see Section \ref{Tevatron-limits}).

If the model under study predicts kinematical distributions
which differ significantly from the distributions of the 
reference processes (\ref{eeHZ-ref}) or (\ref{eeHH-ref}), 
the exclusion result, returned by {\tt HiggsBounds}, can only be 
considered an estimate.\footnote{
In such a case, the general cross section limits   
given by the experimental collaborations 
would have to be replaced by a dedicated
experimental analysis of a particular model giving rise to these 
kinematical distributions, taking into account detailed information about
the detector response to the signal events.
}
\medskip

\subsubsection{LEP input ({HiggsBounds} Input Option {\tt effC}):  effective couplings}
\label{subsubs:LEP-part-IO-effC}
The user is asked to provide model predictions for:
\begin{align}
\nonumber
  & m_{h_k}\,, \;\Gamma_\TOT(h_k)\,,
& & \left(\frac{g^\MOD_{h_kZZ}}{g^\SM_{HZZ}}\right)^2\,,
& & \left(\frac{g^\MOD_{h_k h_i Z}}{g^\REF_{H'HZ}}\right)^2\,, 
& & \left(\frac{g^\MOD_{h_k f\bar f,\EFF}}{g^\SM_{H f\bar f}}\right)^2\,, 
\\
\label{effC-LEP-input}
  & \BR_\MOD(h_k\to h_i h_i) \,,
\end{align}
for $k,i \in \{1,\ldots,n_\HIGGS\}$ and $f \in\{b,\tau\}$.
The normalised effective squared coupling of the Higgs boson
fermion interaction is defined as an approximation for
the ratio of partial decay widths,
\begin{align}
\label{eff-hff-coupling}
\left(\frac{g^\MOD_{h_k f\bar f,\EFF}}{g^\SM_{H f\bar f}}\right)^2
	& := \left.\frac{\Gamma^\MOD_{h_k\to f\bar f}(m_{h_k})}
		   {\Gamma^\SM_{H\to f\bar f}(m_H)}
	\right|_{m_H=m_{h_k}}
\,,
\end{align}
and 
the reference coupling constants are defined by Eq.~(\ref{gref-HHZ})
and by
\begin{align}
\label{gref-HZZ}
(g^\SM_{HZZ})^2 & = 
	\left(\frac{e}{\sw}\frac{m_W}{\cw^2}\right)^2\,,\\
\label{gref-Hff}
(g^\SM_{Hf\bar f})^2 & = 
	\left(\frac{1}{2}\frac{e}{\sw}\frac{m_f}{m_W}\right)^2\,,
\end{align}
where $m_W$ and $m_f$ denote the masses of the $W$~boson 
and fermion $f$, respectively.

For the purpose of running {\tt HiggsBounds}, only the ratio of partial
widths corresponding to the right-hand-side of
Eq.~(\ref{eff-hff-coupling}) is needed. 
In order to completely specify effective couplings for 
neutral Higgs boson interactions 
with fermions, in general a scalar and a pseudoscalar part of the
coupling is needed.
The Feynman rule for the coupling of a generic neutral 
Higgs boson $h$ to fermions can be written as:
\begin{align}
G(h f \bar f) & = i(g_s \text{\bf 1} + i g_p \gamma_5)
\,,
\end{align}
where $g_s$ and $g_p$ are real-valued coupling constants,
and $\text{\bf 1}$ and $\gamma_5$ are the usual matrices in Dirac space.
A scalar particle, like the SM Higgs boson, has $g_p=0$
and a pseudoscalar particle has $g_s=0$.
If the user has the couplings $g_s$ and $g_p$ available, the user may
calculate the normalised effective squared couplings, 
specified in Eq.~(\ref{eff-hff-coupling}), by using the relation:
\begin{align}
\label{eff-hff-coupling2}
\left(\frac{g^\MOD_{h_k f\bar f,\EFF}}{g^\SM_{H f\bar f}}\right)^2
        & = \left(\frac{g^\MOD_{s, h_k f\bar f}}{g^\SM_{H f\bar f}}\right)^2
	+\left(\frac{g^\MOD_{p, h_k f\bar f}}{g^\SM_{H f\bar f}}\right)^2
	 \frac{1}{\beta_f^2(m_{h_k})}
\,,
\end{align}
with 
\begin{align*}
\beta_f(m_{h_k}) & = \sqrt{1-\frac{4 m_f^2}{m_{h_k}^2}}
\,.
\end{align*}
The extra factor of $\beta_f^{-2}$ in Eq.~(\ref{eff-hff-coupling2})
appears, because
the partial width for a scalar
particle decaying into $f \bar f$ is proportional to $\beta_f^3$, while for
a pseudoscalar particle it is proportional to $\beta_f$.

The user may also avoid 
the explicit use of the reference coupling,
defined in Eq.~(\ref{gref-HZZ}),
by using the relation
\begin{align}
\label{ratio-ghzz-from-ratio-gamma}
\left(\frac{g^\MOD_{h_kZZ}}{g^\SM_{HZZ}}\right)^2 & =
	\left.\frac{\Gamma^\MOD_{h_k\to Z^{(\star)} Z^{(\star)}}(m_{h_k})}
		   {\Gamma^\SM_{H\to Z^{(\star)} Z^{(\star)}}(m_H)}
	\right|_{m_H=m_{h_k}}
\,,
\end{align}
where the asterisk in brackets indicates the possible off-shellness
of the decay products. 
Knowledge of the off-shell decay matrix elements
allows to use relation~(\ref{ratio-ghzz-from-ratio-gamma})
for masses $m_{h_k}$ smaller than $2 M_Z$.

From the information specified in Eq.~(\ref{effC-LEP-input}), 
{\tt HiggsBounds} calculates the LEP part of the 
cross section and branching ratios required in 
Input Option {\tt part}
in the effective coupling approximation:
\begin{align}
\frac{\sigma_\MOD(e^+e^-\to h_k Z)}
	{\sigma_\REF(e^+e^-\to H Z)} & =
\left(\frac{g^\MOD_{h_kZZ}}{g^\SM_{HZZ}}\right)^2
\,, \\
\frac{\sigma_\MOD(e^+e^-\to h_k h_i)}
        {\sigma_\REF(e^+e^-\to h_k h_i)} & =
	\left(\frac{g^\MOD_{H'H Z}}{g^\REF_{H'HZ}}\right)^2
\,, \\
\label{BRhff-from-couplings}
\BR_\MOD(h_k \to f\bar f) & = 
	\BR_\SM(H \to f\bar f)(m_H)
	\left.\frac{\Gamma^\SM_\TOT(m_H)}{\Gamma_\TOT(h_k)}
	\right|_{m_H=m_{h_k}}\times \\
\nonumber
& \times \left(\frac{g^\MOD_{h_k f\bar f,\EFF}}{g^\SM_{H f\bar f}}\right)^2
\,,
\end{align}
for $f \in\{b,\tau\}$.

From this point, the evaluation of the quantities $Q_\MOD(X)$
can proceed exactly 
as exemplified in Section \ref{subsubs:LEP-part-IO-part-hadr}.

As far as the implemented LEP constraints are concerned, effective 
couplings just rescale all kinematical distributions of the 
reference processes (\ref{eeHZ-ref}) or (\ref{eeHH-ref})
by common factors. 
Thus, for the class of models where the Higgs sector
can be faithfully parametrised in this way, the 
kinematical distributions correspond to those of the reference
processes. 
Consequently, limits on topological cross sections can exactly be applied.
\medskip

\subsection{Tevatron limits}
\label{Tevatron-limits}

The main SM Higgs boson production processes at the 
proton--anti-proton collider Tevatron at Fermilab are:
\begin{itemize}
\item Gluon fusion: $p\bar p\to H + \cdots$ 
	via $gg\to H$,
\item Higgsstrahlung: $p\bar p\to H V +\cdots$ 
	via $q\bar q'\to V H$ ($V=Z,W^\pm$),
\item Vector boson fusion (VBF): $p\bar p\to H jj + \cdots$ 
	via $q q'\to V^\star V'^\star q'' q''' \to H q'' q'''$,
\item Higgs production associated with heavy quarks:
	$p\bar p\to H t\bar t$ ($H b\bar b$) $+\cdots$,
\end{itemize}
where the ellipses indicate the hadronic remainder of the reaction
which we will suppress in our notation below.

Currently, the Tevatron experiments CDF and D\O\ provide separate limits
on cross sections of individual Higgs search topologies $X$ and 
limits on combinations of search topologies according to SM proportions.
In particular, combinations of Higgs search results 
from {\em both} experiments
are, so far, only available for a SM-like
Higgs boson.
The results implemented in {\tt HiggsBounds} correspond to the Higgs
search topologies that
have been analysed by CDF and D\O, using Tevatron Run II data,
listed in
Tables \ref{TEV-single-toplogies} and \ref{TEV-SM-combined-toplogies}.
New results provided by CDF and D\O\ will be implemented once they
appear, thus keeping {\tt HiggsBounds} up-to-date.

\begin{table}
\begin{tabular}{l|l}
search topology $X$ (analysis)&  reference\\
\hline
$p\bar p \to Z H \to l^+l^- b\bar b$ (CDF with 1.0 fb$^{-1}$) 
	&\cite{CDF-0807-4493}$^{\star}$\\
$p\bar p \to Z H \to l^+l^- b\bar b$ (CDF with 2.4 fb$^{-1}$)		
	&\cite{CDF-9475}\\
$p\bar p \to Z H \to l^+l^- b\bar b$ (D\O\ with 2.3 fb$^{-1}$)
	& \cite{D0-5570}\\
$p\bar p \to W H \to l\nu b\bar b$ (D\O\ with 1.7 fb$^{-1}$)
	& \cite{D0-5472}\\
$p\bar p \to W H \to l\nu b\bar b$ (CDF with 2.7 fb$^{-1}$)
	& \cite{CDF-9463}\\
$p\bar p \to W H \to W^+W^-W^\pm$ (D\O\ with 1.0 fb$^{-1}$)
	& \cite{D0-5485}\\
$p\bar p \to W H \to W^+W^-W^\pm$ (CDF with 1.9 fb$^{-1}$)
	& \cite{CDF-7307}\\
$p\bar p \to H \to W^+W^- \to l^+ l'^-$ (D\O\ with 3.0 fb$^{-1}$)
	& \cite{D0-5757}\\
$p\bar p \to H \to W^+W^- \to l^+ l'^-$ (CDF with 3.0 fb$^{-1}$)
	& \cite{CDF-0809-3930}$^{\star}$\\
$p\bar p \to H \to\gamma\gamma$ (D\O\ with 1.1 fb$^{-1}$)
	& \cite{D0-0803-1514}$^{\star}$\\
$p\bar p \to H \to\gamma\gamma$ (D\O\ with 2.68 fb$^{-1}$)
	& \cite{D0-5737}\\
$p\bar p \to H \to \tau^+\tau^-$ (D\O\ with 1.0 fb$^{-1}$)
	& \cite{D0-0805-2491}$^{\star}$\\
$p\bar p \to H \to \tau^+\tau^-$ (CDF with 1.8 fb$^{-1}$)
	& \cite{CDF-9071}\\
$p\bar p \to b H, H \to b\bar b$ (CDF with 1.9 fb$^{-1}$)
	& \cite{CDF-3b-analysis2}\\
$p\bar p \to b H, H \to b\bar b$ (D\O\ with 1.0 fb$^{-1}$)
	& \cite{D0-0805-3556}$^\star$\\
$p\bar p \to b H, H \to b\bar b$ (D\O\ with 2.6 fb$^{-1}$)
	& \cite{D0-5726}\\
\end{tabular} \\
\caption{\sl\label{TEV-single-toplogies}Tevatron analyses for 
single search topologies, the results
of which are used by {\tt HiggsBounds}. 
The leptons $l$ and $l'$ can be an electron or a muon.
References marked by an asterisk ($\star$) refer to analyses which 
have been published or submitted for publication.
\vspace*{5mm}}
\end{table}

\begin{table}
\begin{tabular}{l|l}
search topologies $\{X\}$ (analysis) & reference\\
\hline
$p\bar p \to W H/Z H \to b\bar b + E_T^{\text{miss.}}$ (CDF with 2.3 fb$^{-1}$)
	& \cite{CDF-9483}\\
$p\bar p \to W H/Z H \to b\bar b + E_T^{\text{miss.}}$ (D\O\ with 2.1 fb$^{-1}$)
	& \cite{D0-5586}\\
$p\bar p \to H/HW/HZ/H \text{\ via VBF}, H\to \tau^+\tau^-$ 
                                                      (CDF with 2.0 fb$^{-1}$) 
	& \cite{CDF-9248}\\
Combined SM analysis (CDF \& D\O\ with 0.9 -- 1.9 fb$^{-1}$) 
        & \cite{CDF-D0-combined-1}\\
Combined SM analysis (CDF \& D\O\ with 1.0 -- 2.4 fb$^{-1}$)  
        & \cite{CDF-D0-combined-2}\\
Combined SM analysis (CDF \& D\O\ with 3.0 fb$^{-1}$)  
        & \cite{CDF-D0-combined-3}\\
\end{tabular}\\
\caption{\sl\label{TEV-SM-combined-toplogies}CDF and D\O\ analyses 
which are contained in {\tt HiggsBounds} and which combine
search topologies according to SM proportions.
\vspace*{5mm}}
\end{table}

In the absence of a Higgs signal, the Tevatron experiments
obtain cross section limits from their measurements:
$\sigma_\MAX(X)$ is the highest cross section for a given assumed
Higgs mass $m_H$ which is compatible with the only-background hypothesis
at 95\% C.L. 
Therefore, if a particular model of the Higgs sector predicts a cross section 
$\sigma_\MOD(X)$ 
for a Higgs boson with mass $m_H$, 
which is higher than the limit, i.e.
\begin{align}
\label{TEV-abs-limit}
\sigma_\MOD(X) & > \sigma_\MAX(X)\,,
\end{align}
then the probability that such a prediction is consistent with observation is 
less than 5\% and the model parameter point will be called excluded 
with (at least) 95\% C.L.
While some Tevatron Higgs search results 
are presented in the form of absolute 
cross section limits, $\sigma_\MAX(X)$,
most of the results are presented as limits
on signal cross sections normalised to the corresponding 
SM cross section, $\sigma_\SM(X)$.\footnote{
So far, no Tevatron Higgs search results have appeared which would
require to use a normalisation to non-SM cross sections.
} 
In the latter case, the exclusion condition~(\ref{TEV-abs-limit}) 
becomes:
\begin{align}
\label{TEV-rel-limit}
\frac{\sigma_\MOD(X)}{\sigma_\SM(X)} 
	& > \frac{\sigma_\MAX(X)}{\sigma_\SM(X)}\,.
\end{align}

In order to test the exclusion 
of a given model by CDF or D\O\ results
according to the procedure outlined 
in Section~\ref{subsec:general approach},
the user needs to provide 
model predictions which allow,
for each relevant Higgs boson search topology $X$,
either absolute cross sections,
\begin{align}
\label{Qmod-abs}
Q_\MOD(X)&= \sigma_\MOD(P(X))\BR_\MOD(F(X))\,,
\end{align}
or cross section ratios,
\begin{align}
\label{Qmod-rel}
Q_\MOD(X)&=\frac{\sigma_\MOD(P(X))\BR_\MOD(F(X))}
	{\sigma_\SM(P(X))\BR_\SM(F(X))}\,,
\end{align}
of hadronic processes to be calculated,
where the symbol $P(X)$ denotes 
the corresponding Higgs production process and $F(X)$
the corresponding Higgs decay chain.
The program allows both variants to be calculated from the 
same input, as the SM predictions for Higgs production processes
and decay branching ratios are internally available.
In Eqs.~(\ref{Qmod-abs}) and (\ref{Qmod-rel}), we have already used 
the implicit assumption that Higgs bosons have narrow width,
for which the CDF and D\O\ limits we employ are valid.
We should also stress again
that the limits are based on analyses which assume 
that the backgrounds to the signal process under study are 
similar to the backgrounds predicted by the SM.
For that reason, the branching ratios for the decays
of the $W^\pm$ or the $Z$ for a particular individual Higgs search topology $X$
are assumed to cancel in the ratio (\ref{Qmod-rel}).

For the calculation of signal cross section predictions
where several Higgs search topologies $\{X_1,\ldots,X_M\}$ are combined
(all cases listed in Table~\ref{TEV-SM-combined-toplogies}),
the definition in Eq.~(\ref{def-Q-combined}) applies.
In the narrow-width approximation, 
this definition
reads:
\begin{align}
\label{Qmod-comb}
Q_\MOD(\{X_1,\ldots X_M\}) &=
        \frac{\sum_{n=1}^{M}\sigma_\MOD(P(X_n))\BR_\MOD(F(X_n))}
             {\sum_{n=1}^{M}\sigma_\SM(P(X_n)\BR_\SM(F(X_n))}
\,.
\end{align}
The quantities $Q_\MOD$ are to be compared with 
the experimental limits $Q_\EXPEC$ and $Q_\OBS$ according to the 
description in Section~\ref{subsec:general approach}. 
There are three options, which allow the user to provide {\tt HiggsBounds}
with the necessary model information 
in order to calculate the quantities $Q_\MOD$.
These are described in the following.
\medskip

\subsubsection{Tevatron input ({HiggsBounds} Input Option {\tt hadr}):
hadronic cross section and branching ratios}
\label{subsubs:Tevatron-part-IO-hadr}
In order to make use of all the implemented Tevatron limits,
the user is asked to provide model predictions for 
the Higgs boson masses $m_{h_k}$ (in units of GeV),
for the branching ratios
\begin{align}
\nonumber
   & \BR_\MOD(h_k\to b\bar b)\,, 
 & & \BR_\MOD(h_k\to \tau^+ \tau^-) \,, \\
\label{tev-hadr-BR}
   & \BR_\MOD(h_k\to W^+ W^-)\,,
&  & \BR_\MOD(h_k\to \gamma\gamma) \,, 
\end{align}
and for the ratios of the hadronic cross sections
\begin{align}
\nonumber
&\frac{\sigma_\MOD(p\bar p \to h_k Z)}
	{\sigma_\SM(p\bar p \to H Z)}\,, 
	& & \frac{\sigma_\MOD(p\bar p \to h_k W^\pm)}
        {\sigma_\SM(p\bar p \to H W^\pm)}\,,
& & \frac{\sigma_\MOD(p\bar p \to h_k \text{ via VBF})}
        {\sigma_\SM(p\bar p \to H \text{ via VBF})}\,,\\
\label{tev-hadr-CS}
&\frac{\sigma_\MOD(p\bar p \to h_k)}
	{\sigma_\SM(p\bar p \to H)}\,, 
	& & \frac{\sigma_\MOD(p\bar p \to h_k b)}
        {\sigma_\SM(p\bar p \to H b)}\,,
\end{align}
for $k\in \{1,\ldots,n_\HIGGS\}$.

With this information and the internally available
SM predictions for Higgs production cross sections and decay branching ratios, 
{\tt HiggsBounds} calculates all values $Q_\MOD$, according to
Eqs.~(\ref{Qmod-abs}), (\ref{Qmod-rel}) or (\ref{Qmod-comb}), which are 
needed in order to test the chosen model against the 
implemented Tevatron results.
Note that the branching ratios $\BR_\MOD(h_k\to h_i h_i)$ are 
currently not used in comparisons with Tevatron results, because, so far, 
no Tevatron analyses have appeared which consider such decays.

While {\tt hadr} is the most widely applicable input option,
it can also be the one that is most difficult for the 
user to provide, as this option requires the 
user to provide predictions for {\em hadronic\/} cross sections,
which involve a convolution with parton distribution functions.
Therefore, the program offers the Input Option {\tt part}, 
which requires only ratios of {\em partonic\/} cross sections.
This should in general be more convenient for the user.
However, this input option is applicable to a narrower class of models
than Input Option {\tt hadr}.

If the model under study predicts kinematical distributions
which differ significantly from the distributions of the 
reference SM processes, 
the exclusion result, returned by {\tt HiggsBounds}, can only be 
considered an estimate 
(as discussed at the end of Section \ref{subsubs:LEP-part-IO-part-hadr}). 
\medskip

\subsubsection{Tevatron input ({HiggsBounds} Input Option {\tt part}):
partonic cross section and branching ratios}
\label{subsubs:Tevatron-part-IO-part}
The user is asked to provide model predictions 
for the Higgs boson masses $m_{h_k}$ (in units of GeV),
for the branching ratios
\begin{align*}
   & \BR_\MOD(h_k\to b\bar b)\,, 
 & & \BR_\MOD(h_k\to \tau^+ \tau^-) \,, \\
   & \BR_\MOD(h_k\to W^+ W^-)\,,
&  & \BR_\MOD(h_k\to \gamma\gamma) \,,
\end{align*}
for the following ratios of partonic cross sections 
\begin{align*}
&\frac{\hat\sigma_\MOD  (g g \to h_k)}
	{\hat\sigma_\SM (g g \to H)}\,,
& &\frac{\hat\sigma_\MOD(b \bar b \to h_k)}
	{\hat\sigma_\SM (b \bar b \to H)}\,, 
& &\frac{\hat\sigma_\MOD(b g \to h_k b)}
	{\hat\sigma_\SM (b g \to H b)}\,, \\
& \frac{\hat\sigma_\MOD (q \bar q' \to h_k W^+)}
        {\hat\sigma_\SM (q \bar q' \to H W^+)}\,,
&& \frac{\hat\sigma_\MOD(q' \bar q \to h_k W^-)}
        {\hat\sigma_\SM (q' \bar q \to H W^-)}\,,
&& (q,q')\in\{(u,d),(c,s)\}\,,\\
& \frac{\hat\sigma_\MOD(q \bar q \to h_k Z)}
        {\hat\sigma_\SM(q \bar q \to H Z)}\,,
&& q\in\{u,d,c,s,b\}\,,
\end{align*}
and for the hadronic cross section ratio
\begin{align*}
& \frac{\sigma_\MOD(p\bar p \to h_k \text{ via VBF})}
        {\sigma_\SM(p\bar p \to H \text{ via VBF})}\,,
\end{align*}
for $k \in \{1,\ldots,n_\HIGGS\}$.

With this information and some internally provided functions (described below) 
{\tt HiggsBounds} can calculate the same input quantities 
as required for Input Option {\tt hadr}
(subject to some approximations, as described below).
It is assumed that the model cross sections for the processes
$\bar b g \to h_k \bar b$ and $b g \to h_k b$ are equal to each other.

Hadronic Higgs production cross sections 
can be written as a sum 
of cross sections of partonic processes $\hat\sigma_{nm\to H+y}(\hat s, m_H)$
convoluted by parton luminosity functions:
\begin{align}
\sigma(p\bar p \to H+y, m_H) & = \int_{\tau_0}^1\!\!d\tau \sum_{\{n,m\}}
	\frac{d{\cal L}_{nm}^{p\bar p}}{d\tau}\,
	\hat\sigma_{nm\to H+y}(\hat s = \tau S, m_H)\,,\\
\nonumber
	& =: \sum_{\{n,m\}} \sigma(p\bar p \to n m \to H+y, m_H)\,.
\end{align}
Here, the sum includes all distinct parton pairs only once.
The centre-of-mass energy of the colliding  proton ($p$) and 
anti-proton ($\bar p$) is assumed to be $\sqrt{S}=1.96\,\tev$.
The symbol $y$ indicates possible further particles in the hard partonic
process, like the $W^\pm$ or $Z$.
The parton luminosity functions are defined as 
\begin{align}
\frac{ d{\cal L}^{AB}_{nm} }{ d\tau } & =
        \int_{\tau}^{1} \frac{dx}{x}
        \frac{1}{1+\delta_{nm}} 
        \Big[
        f_{n/A} (x,\mu_F) f_{m/B} (\frac{\tau}{x},\mu_F)
        + f_{m/A} (x,\mu_F) f_{n/B} (\frac{\tau}{x},\mu_F)
        \Big]
        \, ,
\end{align}
where $f_{n/A} (x,\mu_F)$ denotes the density of partons of type $n$
in the hadron $A$ carrying a fraction $x$ of the hadron's momentum
at the scale $\mu_F$ and $(A,B)=(p,\bar{p})$
for the Tevatron.
The lower bound of the $\tau$-integration ($\tau_0$) is determined
by the minimal invariant mass of the parton system, $\hat s_0 = \tau_0 S$.
The ratio $Q_\MOD(X)$ of Eq.~(\ref{Qmod-rel}) 
factorises into a ratio of branching ratio predictions 
and a ratio of hadronic production cross section predictions.
The calculation of the ratio of branching ratios is straightforward.
In the following, we show an approximate 
relation (that in special cases is exact), which allows to
calculate the ratios of hadronic cross sections
for the Higgs production process $P$,
\begin{align}
\label{R-sigma}
R_\sigma(P) & := \frac{\sigma_\MOD(P)}{\sigma_\SM(P)}
\,,
\end{align}
from the knowledge of ratios of partonic cross sections.
Under the assumption that the ratio of the partonic cross sections,
\begin{align}
\label{R-nm-partonic}
R_{nm}^{H+y}(\hat s, m_H)&:=
  \frac{\hat\sigma^\MOD_{nm\to H+y}(\hat s, m_H)}
       {\hat\sigma^\SM_{nm\to H+y}(\hat s, m_H)} \,,
\end{align}
has, at most, a mild dependence on the parton-system 
centre-of-mass energy squared,
$\hat s$, we can write Eq.~(\ref{R-sigma}) as
\begin{align}
\label{R-sigma-approx}
R_\sigma(P) & \approx \sum_{\{n,m\}}
	R_{nm}^{H+y}(\hat s_0, m_H) 
	\frac{
	  \sigma_\SM(p\bar p \to n m \to H+y, m_H)
	}{\sigma_\SM(p\bar p \to H+y, m_H)}\,,
\end{align}
with $\hat s_0$ denoting the partonic production threshold,
$\hat s_0 =  (m_H+m_y)^2$,
with $m_y=0$ in the case of single Higgs boson production.
In this formula, only ratios of SM hadronic cross sections 
are needed, which we provide in our program. 

\begin{figure}[t]
\centerline{\includegraphics{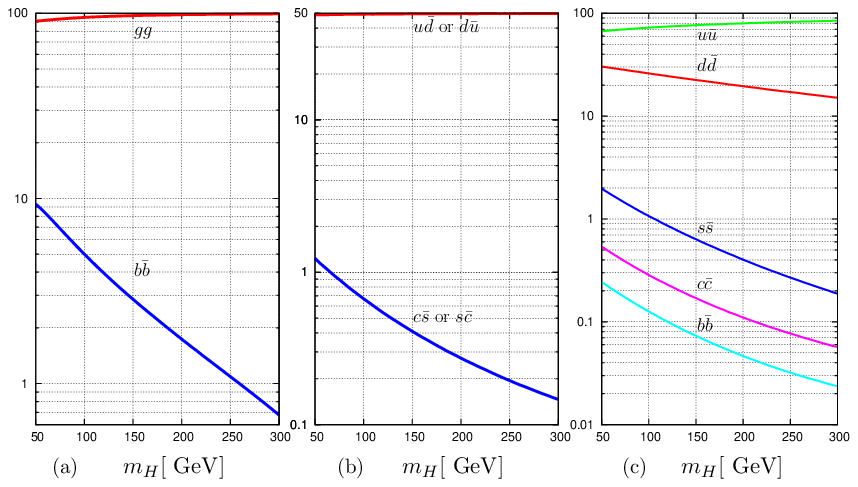}}
    \caption{\sl
	\label{fig:CS-ratios} Relative contributions (in \%) of 
	different SM partonic processes to cross sections of 
	hadronic processes:
	(a) single Higgs production, (b) $HW^\pm$ production
	and (c) $HZ$ production.
        }
\end{figure}

In order to provide the cross section ratios in Eq.~(\ref{R-sigma-approx})
with state-of-the-art theoretical accuracy, we pursue the following strategy. 
The K-factors (NNLO QCD + NLO EW) 
for the $WH$ and $ZH$ production cross sections 
at the Tevatron are, to a very good approximation, independent of the 
generation of the initial state quarks and thus drop out in the ratios. 
The only non-trivial case is the single Higgs production which has 
contributions from gluon fusion and $b \bar b$ annihilation. For the calculation 
of the gluon fusion cross section, we assume a constant NNLO QCD K-factor 
of 3.6 \cite{SM-gluon-fusion-CS-Catani-etal} in the 
relevant range of $70 \gev < m_H < 300 \gev$. 
In this range, our assumption
is consistent with the $m_H$-dependent K-factor reported in 
\cite{SM-gluon-fusion-CS-Catani-etal}
within the uncertainty estimate also given in
\cite{SM-gluon-fusion-CS-Catani-etal}.
In order to obtain an approximate NNLO QCD cross section prediction for the 
b b-bar annihilation process, we used the fact that this cross section
is known to be well
approximated by the LO prediction in the MS-bar scheme with the
factorisation and renormalisation scale choice of $m_H/4$ 
\cite{Harlander:2003ai}.
The SM cross section ratios from Eq.~(\ref{R-sigma-approx}) are then obtained 
by normalising the cross sections for one parton configuration
to the appropriate sum of cross sections.
As a cross-check, we confirmed that our results for Higgs production cross
sections are in good agreement with the compilation of results by 
the TEV4LHC working group \cite{TEV4LHCWG-Higgs-CS}.
Fig.~\ref{fig:CS-ratios} shows our result: the relative contributions
from different parton configurations to the total hadronic cross section
for single Higgs production, $HW^\pm$ production
and $HZ$ production in the range relevant for the 
implemented Tevatron analyses.

Relation (\ref{R-sigma-approx}) becomes an exact identity for all models which 
differ from the SM only by having different values for 
coupling constants, i.e. where the ratios $R_{nm}^{H+y}$ become 
$\hat s$-independent.
Moreover, for models which give $\hat s$-dependent ratios $R_{nm}^{H+y}$,
Eq.~(\ref{R-sigma-approx}) can still be a good approximation because
the luminosity functions $\frac{d{\cal L}_{nm}^{p\bar p}}{d\tau}$
are steeply falling functions of $\tau$ 
and therefore strongly favour the 
threshold regions of the convolution integrals.
However, in this case, the user should make sure that Eq.~(\ref{R-sigma-approx})
is indeed a good approximation for the model under study before using it.
In case of doubt, one can always resort to Input Option {\tt hadr}.

As an example, we show here how the normalised model 
cross section $Q_\MOD(X)$ is obtained on the basis of 
\refeq{R-sigma-approx} for the search topology 
$p\bar p\to H \to W^+ W^-\to l^+l'^-$
of Table~\ref{TEV-single-toplogies} for a model Higgs boson $h_k$:
\begin{multline*}
Q_\MOD(X)= \bigg\{
	\bigg(
		\frac{\hat\sigma^\MOD_{g g \to h_k}(\hat s_0,m_{h_k})}
		     {\hat\sigma^\SM_{g g \to h_k}(\hat s_0,m_H)}
	\; \frac{\sigma_\SM(p\bar p \to g g \to H, m_H)}
		     {\sigma_\SM(p\bar p \to H, m_H)}\\
	+ \frac{\hat\sigma^\MOD_{b\bar b \to h_k}(\hat s_0,m_{h_k})}
                     {\hat\sigma^\SM_{b\bar b \to h_k}(\hat s_0,m_H)}
	\; \frac{\sigma_\SM(p\bar p \to b\bar b \to H, m_H)}
		     {\sigma_\SM(p\bar p \to H, m_H)} 
	\bigg)\times\\
	\times\; 
	\frac{\BR^\MOD_{h_k\to W^+ W^-}(m_{h_k})}{\BR^\SM_{H\to W^+ W^-}(m_H)}
	\bigg\} \bigg|_{m_H=m_{h_k}}
\,.
\end{multline*}

If the model under study predicts kinematical distributions
which differ significantly from the distributions of the 
reference SM processes,
the exclusion result, returned by {\tt HiggsBounds}, can only be 
considered an estimate. 
Further comments regarding this limitation from the end of Section 
\ref{subsubs:LEP-part-IO-part-hadr} apply here too.
\medskip

\subsubsection{Tevatron input ({HiggsBounds} Input Option {\tt effC}):
	effective couplings}
\label{subsubs:Tevatron-part-IO-effC}
%
The user is asked to provide model predictions 
for the Higgs boson masses $m_{h_k}$ 
and total decay widths $\Gamma_\TOT(h_k)$ (in units of GeV),
and for the following normalised effective couplings squared:
\begin{align}
\label{effC-Tevatron-input}
  & \left(\frac{g^\MOD_{h_kgg}}{g^\SM_{Hgg}}\right)^2\,,
& & \left(\frac{g^\MOD_{h_k\gamma\gamma}}{g^\SM_{H\gamma\gamma}}\right)^2\,, 
& & \left(\frac{g^\MOD_{h_kZZ}}{g^\SM_{HZZ}}\right)^2\,, 
& & \left(\frac{g^\MOD_{h_kWW}}{g^\SM_{HWW}}\right)^2\,,
& & \left(\frac{g^\MOD_{h_k f\bar f,\EFF}}{g^\SM_{H f\bar f}}\right)^2\,, 
\end{align}
with $k\in \{1,\ldots,n_\HIGGS\}$ and $f \in\{b,\tau\}$.
The reference coupling constants $g^\SM_{HZZ}$
and $g^\SM_{H f\bar f}$ are defined in Eqs.~(\ref{gref-HZZ}) 
and (\ref{gref-Hff}), respectively, and $g^\SM_{HWW}$ by
\begin{align}
\label{gref-HWW}
(g^\SM_{HWW})^2 & = 
	\left(\frac{e}{\sw}\: m_W\right)^2
\,.
\end{align}
A convenient way to evaluate the ratios of the loop-induced
effective couplings might be to use the ratio of partial decay widths:
\begin{align}
\label{partial-width-ratio}
\left(\frac{g^\MOD_{h_k v v'}}{g^\SM_{H v v'}}\right)^2 & =
	\left.\frac{\Gamma^\MOD_{h_k\to vv'}(m_{h_k})}
		   {\Gamma^\SM_{H\to vv'}(m_H)}
	\right|_{m_H=m_{h_k}}
\,,
\end{align}
with $vv'\in\{ gg,\gamma\gamma \}$.
Of course, the user may apply Eq.~(\ref{partial-width-ratio})
also for $vv'\in\{ ZZ, WW \}$ if $m_H$ is large enough or 
the $\Gamma$s for off-shell vector bosons are used.
If the user has only the 
absolute partial decay widths, $\Gamma^\MOD_{h_k\to F}(m_{h_k})$, available, 
he or she can use the 
built-in functions for the calculation of SM quantities
(see Table~\ref{SMfunctions} below for details) in order to
calculate the right-hand side of Eq.~(\ref{partial-width-ratio}).

From the input specified in Eqs.~(\ref{effC-LEP-input})
and~(\ref{effC-Tevatron-input}) {\tt HiggsBounds} evaluates
all partonic cross section ratios, as required in the Tevatron
part of Input Option {\tt part}, using the following formulae
which are valid in the effective coupling approximation:
\begin{align}
\frac{\hat\sigma_\MOD  (g g \to h_k)}
	{\hat\sigma_\SM (g g \to H)} 
	&=
	\left(\frac{g^\MOD_{h_kgg}}{g^\SM_{Hgg}}\right)^2
\,,\\
\frac{\hat\sigma_\MOD(b \bar b \to h_k)}
	{\hat\sigma_\SM (b \bar b \to H)}
	=
	\frac{\hat\sigma_\MOD(b g \to h_k b)}
        {\hat\sigma_\SM (b g \to H b)}
	&= 
	\left(\frac{g^\MOD_{h_k b\bar b,\EFF}}{g^\SM_{H b\bar b}}\right)^2
\,, \\
\frac{\hat\sigma_\MOD (q \bar q' \to h_k W^+)}
	{\hat\sigma_\SM (q \bar q' \to H W^+)}
	= 
	\frac{\hat\sigma_\MOD(q' \bar q \to h_k W^-)}
        {\hat\sigma_\SM (q' \bar q \to H W^-)} 
	&=
	\left(\frac{g^\MOD_{h_kWW}}{g^\SM_{HWW}}\right)^2
\,,\\
\frac{\hat\sigma_\MOD(q'' \bar q'' \to h_k Z)}
        {\hat\sigma_\SM(q'' \bar q'' \to H Z)}
	&=
	\left(\frac{g^\MOD_{h_kZZ}}{g^\SM_{HZZ}}\right)^2
\,,
\end{align}
for $(q,q')\in\{(u,d),(c,s)\}$ and $q''\in\{u,d,c,s,b\}$.

In order to facilitate the calculation of 
the ratios of the hadronic cross sections 
for $h_k$ production via VBF, we obtained the following 
SM cross section ratio for $p\bar p$ collisions with
1.96 TeV centre-of-mass energy using VBFNLO \cite{VBFatNLO}:
\begin{align}
\label{ratio-SM-WW-vs-ZZ}
\frac{\sigma_\SM(p\bar p \to H \text{ via $WW$ fusion})}
        {\sigma_\SM(p\bar p \to H \text{ via $ZZ$ fusion})}
	& \approx 3.35
\,.
\end{align}
The right-hand side is a mean value. The actual value varies slightly with 
$m_H$ ($\pm 3.6\%$ in the relevant mass range of 
$70\,\gev < m_H < 300\,\gev $). 
Using this mean and neglecting interference effects, 
which affect the result below 1\% (cf. \cite{Djouadi-anatomyI}), 
we can calculate the proportions of $WW$- and $ZZ$-fusion
contributions to the SM VBF Higgs production cross section
for $p\bar p$ collisions with 1.96 TeV centre-of-mass energy:
\begin{align}
R_{\text{VBF}}^{WW} & :=\frac{\sigma_\SM(p\bar p \to H \text{ via $WW$ fusion})}
        {\sigma_\SM(p\bar p \to H \text{ via VBF})}
	= 77\% 
\,,\\
R_{\text{VBF}}^{ZZ} &:= \frac{\sigma_\SM(p\bar p \to H \text{ via $ZZ$ fusion})}
        {\sigma_\SM(p\bar p \to H \text{ via VBF})}
	= 23\%
\,.
\end{align}
The slight variation of the right-hand side of 
Eq.~(\ref{ratio-SM-WW-vs-ZZ}) with $m_H$ 
causes a variation of the above proportions by less than 1\%.

Using the above SM predictions, we can calculate the model predictions
for the ratios of the hadronic cross sections
for $h_k$ production via VBF quite accurately, as follows:
\begin{align}
\frac{\sigma_\MOD(p\bar p \to h_k \text{ via VBF})}
        {\sigma_\SM(p\bar p \to H \text{ via VBF})}
	&=
 	 R_{\text{VBF}}^{WW}\,
		\left(\frac{g^\MOD_{h_kWW}}{g^\SM_{HWW}}\right)^2
	+R_{\text{VBF}}^{ZZ}\,
		\left(\frac{g^\MOD_{h_kZZ}}{g^\SM_{HZZ}}\right)^2
\,.
\end{align}

Branching ratios, additional to the ones we have already described
when discussing the 
LEP limits in Section~\ref{subsubs:LEP-part-IO-effC}, 
are evaluated along the lines of Eq.~(\ref{BRhff-from-couplings}):
\begin{align}
\nonumber
\BR_\MOD(h_k \to \gamma\gamma) & = 
	\BR_\SM(H \to\gamma\gamma )(m_H)
	\left.\frac{\Gamma^\SM_\TOT(m_H)}{\Gamma_\TOT(h_k)}
	\right|_{m_H=m_{h_k}}\times\\
        & \times
	\left(\frac{g^\MOD_{h_k \gamma\gamma}}{g^\SM_{H\gamma\gamma }}\right)^2
\,,\\
\nonumber
\BR_\MOD(h_k \to WW) & = 
	\BR_\SM(H \to WW)(m_H)
	\left.\frac{\Gamma^\SM_\TOT(m_H)}{\Gamma_\TOT(h_k)}
	\right|_{m_H=m_{h_k}}\times\\
	& \times 
	\left(\frac{g^\MOD_{h_k WW}}{g^\SM_{H WW}}\right)^2
\,.
\end{align}
At this point, all quantities which are required as Tevatron input 
by Input Option {\tt part}
are evaluated using the effective coupling approximation.
The evaluation of the model predictions $Q_\MOD(X)$ can
then proceed as exemplified 
in Section \ref{subsubs:Tevatron-part-IO-part} above.

For search topologies, of which the amplitude is proportional to
one monomial of Higgs coupling constants, effective couplings 
just rescale all kinematical distributions of the 
reference  process by a common factor without changing their shape. 
Consequently, in such cases, limits on topological cross sections 
can be exactly applied.
For the application of limits obtained from SM analyses, 
which combine a set of search topologies, the SM-likeness check
(see Section \ref{subsec:general approach}) ensures that
the kinematical distributions arising from models that pass this check
closely resemble the distributions of the SM.
Thus, also in this case, limits on topological cross sections 
can be exactly applied.

Two slightly less straight-forward cases require further discussion.\\[2mm]
a) For the single Higgs production topology, we add the gluon-fusion
and $b\bar b$ annihilation contribution together.
Firstly, one might expect that the boost distribution 
of the produced Higgs boson is different for the two partonic 
processes and thus the kinematical distributions of the Higgs decay products 
would depend on the relative contributions of the two processes 
to the signal. However, it turns out that the gluon and bottom
parton distributions (using MSTW 2008 PDFs \cite{MSTW2008}) 
in the relevant region (lower $x$ values) are, to a very good 
approximation, proportional to each other, i.e.\ the two parton processes 
do not give rise to different boost distributions of the produced 
Higgs boson.
Secondly, it is known that the event selection cuts used in 
single Higgs analyses select Higgs bosons with a certain
$p_T$, which is known to lead to different distributions for
processes arising from gluon-gluon and $b\bar b$ initial states.
This change in the kinematical distribution becomes relevant if the 
coupling of the Higgs to a bottom quark is strongly enhanced such that
the $b\bar b$ annihilation competes with gluon-fusion contribution,   
where the latter dominates in the SM. 
Indeed, it turns out that, for such a scenario,
out of the implemented set of analyses, the most sensitive are
those that involve the search topology $p\bar p\to H\to\tau\tau$
\cite{D0-0805-2491,CDF-9071}. However, these analyses 
reject events with extra jets of large transverse 
momentum, where the deviation of 
the $p_T$-distributions becomes substantial~\cite{langenegger-etal}.     
Therefore, we do not expect a strong influence of the moderately changed
kinematical distributions on the cross section limits. This expectation
is supported by the fact that we reproduce the MSSM exclusion plots in 
\cite{D0-0805-2491,CDF-9071} reasonably well.\\[2mm]
b) For Higgs production via vector boson fusion, a deviation from 
the SM proportion of $WW$ and $ZZ$ initiated processes caused by 
different effective couplings $g_{h_k Z Z}^2$ and $g_{h_k W W}^2$,
does not lead to significant changes in the kinematical distributions
compared to the SM reference process.

In summary, for the implemented set of analyses, we expect the limits
on topological cross sections to be applicable with high accuracy
for models where the Higgs sector can be faithfully parametrised with 
effective couplings.

\section{{\tt HiggsBounds} Operating Instructions}
\label{sec:usage}

There are three formats in which the program {\tt HiggsBounds} can be used:

\begin{itemize}
\item {Library of subroutines}
\item {Command-line version} 
\item {Online version}
\end{itemize}

The most widely applicable format of {\tt HiggsBounds} is the
command-line version, since this reads all the model data from text
files and thus this model data can be generated using any package the
user wishes. The library of {\tt HiggsBounds} subroutines
allows {\tt HiggsBounds} to be 
called within other programs.
If
the user just wishes
to check a few parameter points, the online version provides
quick access to all the functionality of {\tt HiggsBounds}, without the
need to 
install 
the code. 

The {\tt HiggsBounds} code, the online version and documentation can all
be found at the URL {\tt www.ippp.dur.ac.uk/HiggsBounds} .

The {\tt HiggsBounds} code is provided in either Fortran 77 or Fortran 90. 
Both codes provide exactly the same functionality and have exactly the
same operating instructions. 
In fact, the Fortran 77 and Fortran 90 versions of the {\tt HiggsBounds}
subroutines can even be called within codes written in Fortran 90 
and Fortran 77, respectively.
Both codes have also been tested with a variety of Fortran
compilers, including the free gnu compilers which accompany most Linux
distributions. Therefore, the user may download either code and the
difference will only be apparent if the user wishes to examine the
structure of the code\footnote{The maintenance of two separate codes is
  primarily intended to provide an efficient way for the authors to
  confirm that each update is free from implementation errors.}. 

The library of subroutines, the command-line version and the online
version share a common set of features, which we will describe first. We
will then give operating instructions for each of these three 
{\tt HiggsBounds} formats individually. 

\subsection{Common features: Input}

{\tt HiggsBounds} requires four types of input:

\begin{itemize}
\item the number of neutral Higgs bosons in the model under study ({\tt nH}) 
\item the set of experimental analyses which should be considered
  ({\tt whichexpt}) 
\item the theoretical predictions of the model under study (a set of
  input arrays) 
\item the format of these theoretical predictions ({\tt whichinput})
\end{itemize}

Table \ref{instructionsA} contains further information on the variable
{\tt nH}, and the possible values of {\tt whichexpt} are described in
Table \ref{instructions1}. Note that the option {\tt whichexpt=`singH'}
should only be used if neither processes involving Higgs pair production
at LEP nor processes involving the $h_j\to h_ih_i$ decay are
relevant. However, if these conditions are met, this option can save
significantly on computing time. 

{\tt HiggsBounds} expects the theoretical input to be in one of three
formats, labelled by the variable {\tt whichinput}. These formats were
described in detail in Section \ref{sec:implementation} and
are briefly summarised in Table \ref{instructions2}. In
Table \ref{instructions3} and Table \ref{instructions3a} we assign names
to all of the possible input arrays (each array is defined in terms of
the notation used in Section \ref{sec:implementation}). These names will
prove useful when we describe the input requirements of each version of
{\tt HiggsBounds} individually. 

\begin{table}[!h]
\begin{tabular}{ll}
{\tt nH} & ({\tt integer}) \\
\hline
1-9& By default, {\tt HiggsBounds} can be applied to models with \\[-2mm]
& between 1 and 9 neutral Higgs bosons. \\[-2mm]
& (This range can easily be extended by the user if required.)
\end{tabular}\\
\caption{\sl\label{instructionsA}The possible values of the variable 
  {\tt nH}, which labels the number of neutral Higgs bosons 
in the model under
  study.
} 
\end{table}

\begin{table}[!h]
\begin{tabular}{ll}
{\tt whichexpt} & ({\tt character(LEN=5)}) \\
\hline
{\tt LandT} & both LEP and Tevatron analyses \\
{\tt onlyL} & only LEP analyses\\
{\tt onlyT} & only Tevatron analyses\\
{\tt singH} & only analyses for processes involving one Higgs boson\\
\end{tabular}\\
\caption{\sl\label{instructions1}The possible values of the variable {\tt
    whichexpt}, which indicates which subset of experimental analyses
  will be considered by {\tt HiggsBounds}.
} 
\end{table}

\begin{table}[!h]
\begin{tabular}{ll}
{\tt whichinput} & ({\tt character(LEN=4)}) \\
\hline
{\tt effC} & Masses, total decay widths, \\[-2mm]
           & ratios of effective couplings squared, ${\rm BR}(h_j\to h_ih_i)$.\\
{\tt part} & Masses, total decay widths, ratios of LEP cross sections, \\[-2mm]
           & mainly ratios of partonic Tevatron cross sections, branching ratios.\\
{\tt hadr} & Masses, total decay widths, ratios of LEP cross sections, \\[-2mm]
           & ratios of hadronic Tevatron cross sections, branching ratios.\\
\end{tabular}\\
\caption{\sl\label{instructions2}The possible values of the variable {\tt
    whichinput}, which indicates the format of the theoretical
  predictions provided by the user. (See Section
  \ref{sec:implementation} for a more detailed description of each of
  these settings).} 
\end{table}

\clearpage

\begin{table}[h]
\begin{tabular}{llll}
input arrays &  \multicolumn{3}{l}{\tt (double precision)}\\
\hline
{\tt Mh(nH)}         &  $m_{h_i}$ &               \multicolumn{2}{l}{in GeV}\\
{\tt GammaTotal(nH)} &  $\Gamma_{\rm tot}(h_i)$ & \multicolumn{2}{l}{in GeV} \\
\hline
{\tt g2hjbb(nH)}      & $\left(\frac{g^{\rm model}_{h_j{\rm (OP)}}}
                                    {g^{\rm SM}_{H{\rm (OP)}}}\right)^2$,& OP =  
                          & $b\bar{b}$     \\
{\tt g2hjtautau(nH)}  & & & $\tau^+\tau^-$ \\
{\tt g2hjWW(nH)}      & & & $WW$           \\
{\tt g2hjZZ(nH)}      & & & $ZZ$           \\
{\tt g2hjgaga(nH)}    & & & $\gamma\gamma$ \\
{\tt g2hjgg(nH)}      & & & $gg$           \\
\hline
{\tt g2hjhiZ(nH,nH)} & $\left(\frac{g^{\rm model}_{h_jh_iZ}}
                                   {g^{\rm ref}_{HH^{\prime}Z}}\right)^2$ &&\\
\hline
{\tt BR\_hjbb(nH)}      & BR($h_j \to {\rm OP}$),& OP = &$b\bar{b}$     \\
{\tt BR\_hjtautau(nH)}& &                               &$\tau^+\tau^-$ \\
{\tt BR\_hjWW(nH)}    & &                               &$WW$           \\
{\tt BR\_hjgaga(nH)}  & &                               &$\gamma\gamma$ \\
\hline
{\tt BR\_hjhihi(nH,nH)} & \multicolumn{3}{l}{BR($h_j\to h_ih_i$)} \\
\hline
\end{tabular}\\
\caption{\sl\label{instructions3}Input arrays for model predictions for 
	 effective normalised squared couplings and branching ratios
    recognised by {\tt HiggsBounds}.
  The size of
  each array is given in brackets in the first column. See Section
  \ref{sec:implementation} for the description of the notation used
  in the second column.
  The elements of {\tt BR\_hjhihi} are ordered such that 
  {\tt BR\_hjhihi(j,i)}$ = \BR(h_j\to h_ih_i)$. 
} 
\end{table}

\begin{table}[!h]
\begin{tabular}{llll}
input arrays cont. & \multicolumn{3}{l}{\tt (double precision)}\\
\hline 
{\tt CS\_lep\_hjZ\_ratio(nH)}             & $R_{\sigma}(P)$,
                                          & $P$ =
                                            & $e^+e^- \to h_j Z$   \\
{\tt CS\_lep\_hjhi\_ratio(nH,nH)}         & & & $e^+e^- \to h_j h_i$ \\
{\tt CS\_tev\_pp\_hj\_ratio(nH)}           & & & $p\bar{p} \to h_j$   \\
{\tt CS\_tev\_pp\_hjb\_ratio(nH)}          & & & $p\bar{p} \to b h_j$ \\
{\tt CS\_tev\_pp\_hjW\_ratio(nH)}          & & & $p\bar{p} \to h_j W$ \\
{\tt CS\_tev\_pp\_hjZ\_ratio(nH)}          & & & $p\bar{p} \to h_j Z$\\
{\tt CS\_tev\_pp\_vbf\_ratio(nH)}          & & & $p\bar{p} \to h_j 
                                                 {\rm \, via \, VBF}$\\
\hline
{\tt CS\_tev\_gg\_hj\_ratio(nH)}           & $R^{h_j}_{nm}$,
                                          & $nm$ = 
                                            & $gg$ \\
{\tt CS\_tev\_bb\_hj\_ratio(nH)}           & & & $b\bar{b}$   \\
\hline
{\tt CS\_tev\_ud\_hjWp\_ratio(nH)}         & $R^{h_j +W^+}_{nm}$,
                                          & $nm$ = 
                                            & $u\bar{d}$ \\
{\tt CS\_tev\_cs\_hjWp\_ratio(nH)}         & & & $c\bar{s}$ \\
\hline
{\tt CS\_tev\_ud\_hjWm\_ratio(nH)}         &$R^{h_j +W^-}_{nm}$
                                          & $nm$ =
                                            &  $d\bar{u}$ \\
{\tt CS\_tev\_cs\_hjWm\_ratio(nH)}         & & &  $s\bar{c}$ \\
\hline
{\tt CS\_tev\_dd\_hjZ\_ratio(nH)}          &$R^{h_j +Z}_{nm}$
                                          & $nm$ =                
                                            & $d\bar{d}$ \\
{\tt CS\_tev\_uu\_hjZ\_ratio(nH)}          & & & $u\bar{u}$ \\
{\tt CS\_tev\_ss\_hjZ\_ratio(nH)}          & & & $s\bar{s}$ \\
{\tt CS\_tev\_cc\_hjZ\_ratio(nH)}          & & & $c\bar{c}$ \\
{\tt CS\_tev\_bb\_hjZ\_ratio(nH)}          & & & $b\bar{b}$ \\
\hline
{\tt CS\_tev\_bg\_hjb\_ratio(nH)}          &$R^{h_j +b}_{nm}$
                                          & $nm$ = 
                                            &  $bg$, $\bar{b}g$ \\
\hline
\end{tabular}\\
\caption{\sl\label{instructions3a}Input arrays 
	for model predictions for cross section ratios
	recognised by {\tt HiggsBounds}.
  The size of
  each array is given in brackets in the first column. The LEP or
  hadronic Tevatron cross section ratios $R_{\sigma}(P)$ are defined in
  Eq.~(\ref{R-sigma}) and the partonic Tevatron cross section ratios
  $R^{h_j+y}_{nm}$ are defined in Eq.~(\ref{R-nm-partonic}).
\vspace*{5mm}} 
\end{table}

\clearpage

\subsection{Common features: Output}
\label{subsec:Common features: Output}

{\tt HiggsBounds} provides the user with four types of output:

\begin{itemize}
\item whether the parameter point is excluded at the 95\% C.L. or not
  ({\tt HBresult}) 
\item the identifying number of the process with the highest statistical
  sensitivity ({\tt chan}).  
\item the number of Higgs bosons which have contributed to the theoretical rate
  for this process ({\tt ncombined}) 
\item the ratio of the theoretical rate $Q_{\rm model}$ to the observed
  limit $Q_{\rm obs}$ for this process ({\tt obsratio}). 
\end{itemize}

Table \ref{instructions4} shows the possible values of {\tt HBresult}
and {\tt obsratio}, which are complementary. Tables \ref{instructions4a}
and \ref{instructions4b} discuss {\tt chan} and {\tt ncombined}
respectively. If the library of subroutines or the command-line versions
are used, the key to the process numbers is written in the file {\tt
  Key.dat}. In the online version, this information appears on the
screen.  

\begin{table}[!h]
\begin{tabular}{lll}
\hline
{\tt HBresult} &{\tt obsratio}&\\
 ({\tt integer})&({\tt double precision})&\\
\hline
           0 &   $\ge 1.0$   & parameter point is excluded\\
           1 &   $<1.0$      & parameter point is not excluded\\
          -1 &   $\le 0.0$   & invalid parameter set\\
\end{tabular}
\caption{\sl\label{instructions4}The possible values of the output variables {\tt
    HBresult} and {\tt obsratio}, which indicate whether a parameter
  point has been excluded at the 95\% C.L. by the experimental results
  under consideration.
\vspace*{5mm}} 
\end{table} 

\begin{table}[!h]
\begin{tabular}{ll}
{\tt chan} & ({\tt integer})\\
\hline
1-[\# of considered analyses]
 & See the file {\tt Key.dat} for the definition of each  \\[-2mm]
 & process number. {\tt Key.dat} is automatically\\ [-2mm]
 & generated when either the command line or the\\[-2mm]
 & subroutine version of {\tt HiggsBounds} are used.
\end{tabular}\\
\caption{\sl\label{instructions4a}Further information about the output variable
  {\tt chan}, which stores the reference number of the process with the
  highest statistical sensitivity.
\vspace*{5mm}} 
\end{table} 

\begin{table}[!h]
\begin{tabular}{ll}
{\tt ncombined}& ({\tt integer})\\
\hline
1-$n_H$& Number of Higgs bosons which have contributed to the theoretical\\[-2mm] 
& rate for this process. The number depends on {\tt delta\_Mh\_LEP}\\[-2mm]
& or {\tt delta\_Mh\_TEV}.
\end{tabular}\\
\caption{\sl\label{instructions4b}Further information 
about the output variable {\tt ncombined}.
\vspace*{5mm}}
\end{table} 

\newpage

\subsection{Library of subroutines}
\label{subsec:Library of subroutines}

{\it Installation}\\[.3cm]
The {\tt HiggsBounds} code can be compiled to form a library of
subroutines using the following commands: 

\begin{verbatim}
./configure
make libHB
\end{verbatim}

A program which wishes to use the {\tt HiggsBounds} subroutines can be
compiled and linked to the library by adding {\tt -L<HBpath> -lHB} to
the command line, for example, 

\begin{verbatim}
gfortran myprog.f90 -o myprog -L<HBpath> -lHB
\end{verbatim}

where {\tt <HBpath>} is the location of the {\tt HiggsBounds} library. 
\medskip

{\it Subroutine} {\tt initialize\_HiggsBounds}\\[.3cm]
The subroutine {\tt initialize\_HiggsBounds} must be called before any
other {\tt HiggsBounds} subroutine. It performs some preparatory
operations such as reading in the tables of data. It is called as: 
\begin{verbatim}
call initialize_HiggsBounds(nH, whichexpt)
\end{verbatim}

When using the {\tt HiggsBounds} subroutines in another code, the
subroutine {\tt initialize\_HiggsBounds} must be called only once,
before any other \linebreak {\tt HiggsBounds} subroutine is called. 
\medskip

{\it Subroutines} {\tt run\_HiggsBounds\_effC, run\_HiggsBounds\_part}
{\it and} \\
  {\tt run\_HiggsBounds\_hadr}\\[.3cm]
The subroutines {\tt run\_HiggsBounds\_effC}, 
{\tt run\_HiggsBounds\_part} and 
\linebreak 
{\tt run\_HiggsBounds\_hadr} perform the main part of the {\tt
HiggsBounds} calculations. They set the value of {\tt whichinput} to be
{\tt effC}, {\tt part} and {\tt hadr} respectively and therefore require
different arguments. These subroutines are called as:
\begin{verbatim}
call run_HiggsBounds_effC(nH,Mh,GammaTotal,
     &          g2hjbb,g2hjtautau,g2hjWW,g2hjZZ,              
     &          g2hjgaga,g2hjgg,g2hjhiZ,                      
     &          BR_hjhihi,                                    
     &          HBresult,chan,                          
     &          obsratio, ncombined                          )
\end{verbatim}

\begin{verbatim}
call run_HiggsBounds_part(nH,Mh,                    
     &          CS_lep_hjZ_ratio,    CS_lep_hjhi_ratio,                            
     &          CS_tev_gg_hj_ratio,  CS_tev_bb_hj_ratio,        
     &          CS_tev_bg_hjb_ratio,                          
     &          CS_tev_ud_hjWp_ratio,CS_tev_cs_hjWp_ratio,     
     &          CS_tev_ud_hjWm_ratio,CS_tev_cs_hjWm_ratio,     
     &          CS_tev_dd_hjZ_ratio, CS_tev_uu_hjZ_ratio,      
     &          CS_tev_ss_hjZ_ratio, CS_tev_cc_hjZ_ratio,      
     &          CS_tev_bb_hjZ_ratio,                          
     &          CS_tev_pp_vbf_ratio,                          
     &          BR_hjbb,BR_hjtautau,                          
     &          BR_hjWW,BR_hjgaga,                             
     &          BR_hjhihi,                                     
     &          HBresult,chan,                          
     &          obsratio, ncombined                          )
\end{verbatim}

\begin{verbatim}
call run_HiggsBounds_hadr(nH,Mh,                    
     &          CS_lep_hjZ_ratio,    CS_lep_hjhi_ratio,           
     &          CS_tev_pp_hj_ratio,  CS_tev_pp_hjb_ratio,      
     &          CS_tev_pp_hjW_ratio, CS_tev_pp_hjZ_ratio,      
     &          CS_tev_pp_vbf_ratio,                          
     &          BR_hjbb,BR_hjtautau,                         
     &          BR_hjWW,BR_hjgaga,                            
     &          BR_hjhihi,                                    
     &          HBresult,chan,                          
     &          obsratio, ncombined                          )
\end{verbatim}

Each of these arguments must be supplied. However, if 
a branching ratio, effective coupling or cross section is believed 
to be irrelevant, the corresponding array may be filled with zeros. 
This will ensure that the value of
$Q_{\rm model}$ for processes involving this quantity will also be
zero. For example, in the MSSM, processes involving the decay $h_j \to
\gamma \gamma$ will rarely be the process with the highest statistical
sensitivity of $Q_{\rm model}/Q_{\rm expec}$ and, consequently, 
it may be convenient to set the arrays
{\tt g2hjgaga} and {\tt BR\_hjgaga} to zero for simplicity. 

Also, depending on the value given for {\tt whichexpt}, some of the
input arrays will be ignored within {\tt HiggsBounds}. For example, if
{\tt whichexpt=`onlyT'}, the branching ratio for the Higgs cascade decay
$h_j \to h_i h_i$
will not be relevant. Therefore, setting this array to zero in this case
will not affect the {\tt HiggsBounds} results in this case. 
\medskip
 
{\it Subroutine} {\tt finish\_HiggsBounds}\\[.3cm]
The subroutine {\tt finish\_HiggsBounds} should be called once at the
end of the program, after all other {\tt HiggsBounds}
subroutines\footnote{In the Fortran 90 version of the code, the
  subroutine {\tt finish\_HiggsBounds} is used to deallocate the
  allocatable arrays used within {\tt HiggsBounds}.}. It is called as: 
 
\begin{verbatim}
call finish_HiggsBounds
\end{verbatim}
\medskip

{\it Functions for Standard Model branching ratios, total decay width
  and cross sections}\\[.3cm]
The {\tt HiggsBounds} library also allows users access to the Standard Model
Higgs branching ratios, total decay width and production cross sections,
which are used internally by {\tt HiggsBounds}. We use Standard Model Higgs
branching ratios and total decay width from the program 
HDECAY 3.303 \cite{hdecay}. 
The SM hadronic cross sections have been obtained from 
the TEV4LHC Higgs Working Group \cite{TEV4LHCWG-Higgs-CS}
(see Table \ref{SMfunctions} for references to the original works)
with the exception of the $\sigma^{\rm SM}(p\bar{p}\to b g \to b H)$
cross section. The latter cross section has been calculated with
the program HJET 1.1 \cite{HJET} for a set of different cuts
on the transverse momentum and pseudo-rapidity of the $b$-quark,
which are needed internally in order to apply correctly the results
of some Tevatron analyses and which were not available from
\cite{TEV4LHCWG-Higgs-CS}. From this set, only the cross section 
without cuts is externally provided.
\medskip

\begin{table}
\begin{tabular}{lllll}
\hline
        function                &\multicolumn{3}{l}{({\tt double precision})} & \\
\hline
      {\tt  SMGamma\_h(Mh)     } &\multicolumn{3}{l}{$\Gamma^{\rm SM}_{\rm tot} (h_i)$}&\cite{hdecay}\\
      {\tt  SMBR\_Hbb(Mh)      } & ${\rm BR}^{\rm SM}$($H \to$ OP), & OP= & $b \bar{b}$&\cite{hdecay}\\
      {\tt  SMBR\_Htautau(Mh)  } &                      &     & $\tau^- \tau^+ $&\cite{hdecay}\\
      {\tt  SMBR\_HWW(Mh)      } &                      &     & $WW            $&\cite{hdecay}\\
      {\tt  SMBR\_Hgamgam(Mh)  } &                      &     & $\gamma \gamma $&\cite{hdecay}\\
      {\tt  SMBR\_Hgg(Mh)      } &                      &     & $gg            $&\cite{hdecay}\\
      {\tt  SMBR\_HZgam(Mh)    } &                      &     & $Z \gamma      $&\cite{hdecay}\\
      {\tt  SMBR\_HZZ(Mh)      } &                      &     & $ZZ            $&\cite{hdecay}\\
      {\tt  SMBR\_Htoptop(Mh)  } &                      &     & $t \bar{t}     $&\cite{hdecay}\\
      {\tt  SMBR\_Hcc(Mh)      } &                      &     & $c \bar{c}     $&\cite{hdecay}\\
      {\tt  SMBR\_Hss(Mh)      } &                      &     & $s \bar{s}     $&\cite{hdecay}\\
      {\tt  SMBR\_Hmumu(Mh)    } &                      &     & $\mu^- \mu^+   $&\cite{hdecay}\\
\hline
      {\tt  SMCS\_tev\_pp\_qq\_HW(Mh)} & $\sigma^{\rm SM}(P)$, & P=  & $p\bar{p}\to q\bar{q}\to HW$&\cite{SM-VH-CS}\\
      {\tt  SMCS\_tev\_pp\_qq\_HZ(Mh)} &                      &     & $p\bar{p}\to q\bar{q}\to HZ$&\cite{SM-VH-CS}\\
      {\tt  SMCS\_tev\_pp\_gg\_H(Mh) } &                      &     & $p\bar{p}\to gg      \to H$&\cite{SM-gluon-fusion-CS,SM-gluon-fusion-CS-Catani-etal}\\
      {\tt  SMCS\_tev\_pp\_bb\_H(Mh) } &                      &     & $p\bar{p}\to b\bar{b}\to H $&\cite{Harlander:2003ai}\\
      {\tt  SMCS\_tev\_pp\_vbf\_H(Mh)} &                      &     & $p\bar{p}\to H $ via VBF &\cite{SM-VBF-CS}\\
      {\tt  SMCS\_tev\_pp\_bg\_Hb(Mh)} &                      &     & $p\bar{p}\to bg\to Hb    $&\cite{HJET}\\
\hline
\end{tabular}\\
\caption{\sl\label{SMfunctions}
  Standard Model branching ratios, total decay widths in units of GeV
  and hadronic Tevatron cross sections in units of pikobarn
  provided as functions by
  {\tt HiggsBounds}, together with references. Each function takes a
  Higgs mass {\tt Mh (double precision)} as its argument.
\vspace*{5mm}} 
\end{table}

{\it Examples}\\[.3cm]
We have provided three example programs which demonstrate the use of the
{\tt HiggsBounds} subroutines. The first example relates to
the Fourth Generation Model and is contained in the file {\tt
   example-SM\_vs\_4thGen.F} (see Fig.~\ref{fig:fourth-gen-code} 
below for a code listing). 
This program uses the {\tt HiggsBounds} functions for
the SM branching ratios and SM total decay width to calculate the Higgs
decay width and the effective normalised squared couplings
in the SM and a simple Fourth Generation Model. 
This information is then used as input for
the subroutine {\tt run\_HiggsBounds\_effC}, which is called
once with SM input and once with Fourth Generation Model input.
The example will be described in more detail in Section \ref{subsec:FGM}. 
Once the {\tt HiggsBounds} library has been compiled
(using {\tt ./configure ; make libHB} as described previously), the code
 {\tt example-SM\_vs\_4thGen.F} can be compiled and run with the commands: 

\begin{verbatim}
gfortran  example-SM_vs_4thGen.F -o HBfourthgen -L<HBpath> -lHB
./HBfourthgen
\end{verbatim}

where {\tt <HBpath>} is the location of the {\tt HiggsBounds} library. 

The files {\tt HBwithFH.F} and {\tt HBwithCPsuperH.f} demonstrate the
use of \linebreak {\tt HiggsBounds} subroutines with the publically
available programs {\tt FeynHiggs}~\cite{feynhiggs} and {\tt
  CPsuperH}~\cite{cpsh}, respectively. We refer the reader to the
extensive comments contained within these example files for further
details. 

\subsection{Command line version}
{\it Installation}\\[.3cm]
In order to be able to call {\tt HiggsBounds} from the command line, it
should be compiled using the commands 
\begin{verbatim}
./configure
make
\end{verbatim}
\medskip

{\it Command line and input file format}\\[.3cm]
In the command-line usage of {\tt HiggsBounds}, the arrays containing
the theoretical model predictions are read from text files. The other
options are specified in the command line, which is of the form: 
\begin{verbatim}
./HiggsBounds <whichexpt> <whichinput> <nH> <prefix>
\end{verbatim}

The variable {\tt <prefix>} is a string which is added to the front of
input and output file names and may include directory names or other
identifying information. 

Table \ref{contentsoffiles} describes the contents of each input
file. Note that each input file should start with a line number.  The
input files should not contain any comments or blank lines. 
The line number identifies the predictions which belong to the same
model parameter point in different files.

\begin{table}[!t]
\begin{tabular}{ll}
\hline
file name & data format\\
\hline
{\tt MH\_GammaTot.dat}                & {\tt k}, {\tt Mh, GammaTotal}\\[-1mm]
{\tt effC.dat}                        & {\tt k}, {\tt g2hjbb,g2hjtautau,g2hjWW,g2hjZZ,} \\[-2mm]
                                      &\phantom{{\tt k},} {\tt g2hjgaga,g2hjgg,} \\[-2mm]
                                      &\phantom{{\tt k},} {some elements of {\tt g2hjhiZ}}\\[-2mm]
                                      &\phantom{{\tt k},} {(lower left triangle - see example)}\\[-1mm]    
{\tt LEP\_HZ\_CS\_ratios.dat}         & {\tt k}, {\tt CS\_lep\_hjZ\_ratio}\\[-1mm]  
{\tt LEP\_2H\_CS\_ratios.dat}         &\phantom{{\tt k},} some elements of {\tt CS\_lep\_hjhi\_ratio} \\[-2mm]
                                      &\phantom{{\tt k},} (lower left triangle - see example)\\[-1mm]
{\tt TEV\_H\_0jet\_partCS\_ratios.dat}& {\tt k}, {\tt CS\_tev\_gg\_hj\_ratio,CS\_tev\_bb\_hj\_ratio}\\[-1mm]   
{\tt TEV\_H\_1jet\_partCS\_ratios.dat}& {\tt k} {\tt CS\_tev\_bb\_hjb\_ratio}\\[-1mm]  
{\tt TEV\_HW\_partCS\_ratios.dat}     & {\tt k}, {\tt CS\_tev\_ud\_hjWp\_ratio},\\[-2mm]  
				      &\phantom{{\tt k},} {\tt CS\_tev\_cs\_hjWp\_ratio},\\[-2mm]  
                                      &\phantom{{\tt k},} {\tt CS\_tev\_ud\_hjWm\_ratio},\\[-2mm]
				      &\phantom{{\tt k},} {\tt CS\_tev\_cs\_hjWm\_ratio}\\[-1mm]  
{\tt TEV\_HZ\_partCS\_ratios.dat}     & {\tt k}, {\tt CS\_tev\_dd\_hjZ\_ratio},\\[-2mm]
				      &\phantom{{\tt k},} {\tt CS\_tev\_uu\_hjZ\_ratio,}\\[-2mm] 
                                      &\phantom{{\tt k},} {\tt CS\_tev\_ss\_hjZ\_ratio},\\[-2mm]
				      &\phantom{{\tt k},} {\tt CS\_tev\_cc\_hjZ\_ratio,}\\[-2mm] 
                                      &\phantom{{\tt k},} {\tt CS\_tev\_bb\_hjZ\_ratio}\\[-1mm]
{\tt TEV\_H\_vbf\_hadCS\_ratios.dat}  & {\tt k}, {\tt CS\_tev\_pp\_vbf\_ratio}\\[-1mm] 
{\tt TEV\_1H\_hadCS\_ratios.dat}      & {\tt k}, {\tt CS\_tev\_pp\_hj\_ratio},\\[-2mm]
				      &\phantom{{\tt k},} {\tt CS\_tev\_pp\_hjb\_ratio},\\[-2mm]
                                      &\phantom{{\tt k},} {\tt CS\_tev\_pp\_hjW\_ratio},\\[-2mm]
				      &\phantom{{\tt k},} {\tt CS\_tev\_pp\_hjZ\_ratio,}\\[-2mm]
                                      &\phantom{{\tt k},} {\tt CS\_tev\_pp\_vbf\_ratio}\\[-1mm] 
{\tt BR\_1H.dat}                      & {\tt k}, {\tt BR\_hjbb,BR\_hjtautau,BR\_hjWW,BR\_hjgaga} \\[-1mm]   
{\tt BR\_2H.dat}                      & {\tt k}, some elements of {\tt BR\_hjhihi} \\[-2mm]
                                      &\phantom{{\tt k},} (row by row, without diagonal \\[-2mm]
                                      &\phantom{{\tt k},} - see example)\\[-1mm]
{\tt additional.dat}(optional)        & {\tt k},  ... \\
\hline
\end{tabular}\\
\caption{\sl\label{contentsoffiles}Names and data format of 
  all {\tt HiggsBounds} input files.
  The right column shows the order of the input data arrays 
  within one line of the input file. For the order
  within the arrays, see example. 
  {\tt k} is the line number. Note that the
  array {\tt CS\_tev\_pp\_vbf\_ratio} appears in two different input
  files. However, these files will never be required by 
  {\tt HiggsBounds} simultaneously.
\vspace*{5mm}} 
\end{table}

Care should be taken with the order of the array elements in the
files. This is best illustrated by an example, where we will use
$n_H=3$. The one dimensional arrays, e.g. {\tt Mh}, should be given in
the order 
\begin{quote}
{\tt Mh(1), Mh(2), Mh(3)}
\end{quote}
However, not all of the elements of the two dimensional arrays are
required. Only the lower left triangle (including the diagonal) is
required from the arrays {\tt g2hjhiZ} and {\tt lepCS\_hjhi\_ratio},
since they are symmetric, e.g.  
\begin{align}
\begin{pmatrix}
{\tt g2hjhiZ(1,1)} & \Gray{\tt g2hjhiZ(1,2)} & \Gray{\tt g2hjhiZ(1,3)}\non\\
{\tt g2hjhiZ(2,1)} &      {\tt g2hjhiZ(2,2)} & \Gray{\tt g2hjhiZ(2,3)}\non\\
{\tt g2hjhiZ(3,1)} &      {\tt g2hjhiZ(3,2)} &      {\tt g2hjhiZ(3,3)}\non
\end{pmatrix}
\end{align}
i.e. the elements in the input file should be written in the order 
\begin{quote}
{\tt g2hjhiZ(1,1)}, {\tt g2hjhiZ(2,1)}, {\tt g2hjhiZ(2,2)}, 
{\tt g2hjhiZ(3,1)}, \\* {\tt g2hjhiZ(3,2)}, {\tt g2hjhiZ(3,3)} .
\end{quote}
\clearpage 

For the array {\tt BR\_hjhihi}, only the off-diagonal components are required
\begin{align}
\begin{pmatrix} 
\Gray{\tt BR\_hjhihi(1,1)} & {\tt BR\_hjhihi(1,2)}     & {\tt BR\_hjhihi(1,3)} 
                                                                         \non\\
{\tt BR\_hjhihi(2,1)}      & \Gray{\tt BR\_hjhihi(2,2)} & {\tt BR\_hjhihi(2,3)} 
                                                                          \non\\
{\tt BR\_hjhihi(3,1)}      & {\tt BR\_hjhihi(3,2)}      & 
                                                \Gray{\tt BR\_hjhihi(3,3)}\non
\end{pmatrix}
\end{align}
since the diagonal elements are not physical quantities. Therefore, the
elements should be written in the order  
\begin{quote}
{\tt BR\_hjhihi(1,2)}, {\tt BR\_hjhihi(1,3)}, {\tt BR\_hjhihi(2,1)},
{\tt BR\_hjhihi(2,3)}, \\* {\tt BR\_hjhihi(3,1)}, {\tt BR\_hjhihi(3,2)}  
\end{quote}
in the input file.

The file {\tt additional.dat} is optional. If it is included, it can
have any number of columns greater than 1 (as for the previous files,
the first entry on each line should be the line number). It is envisaged
that this input file will be particularly useful when parameter scans
are performed over a variable which is not required by {\tt HiggsBounds}
but helpful plotting the results.  For example, in the case of the
complex MSSM, {\tt additional.dat} could be used to store the values of
$\tan \beta$ and the charged Higgs mass. 

As in the subroutine version, the command line version of 
{\tt HiggsBounds} expects a subset of the total list of input arrays, which
depends on the chosen setting of {\tt whichinput}. The maximal list of files used for
each value of {\tt whichinput} is given in Table \ref{instructions1}.  

As discussed for the subroutine version, some of the arrays will not be
relevant for some of the choices for {\tt whichexpt}. The command line
version of {\tt HiggsBounds} will consider the list of input files
appropriate to the setting {\tt whichinput} and then only attempt to
read any of these input files if the value chosen for {\tt whichexpt}
means that at least one of the arrays it contains will be directly
used. Table \ref{instructions6} contains a list of which input files are
actually relevant to each value of {\tt whichexpt}.  
For example, if {\tt whichinput = 'hadr'},  and {\tt whichexpt = 'LandT'},
then {\tt HiggsBounds} requires the input files: 

\begin{quote}
{\tt MH\_GammaTot.dat}, {\tt BR\_2H.dat}, {\tt BR\_1H.dat}, 
{\tt LEP\_HZ\_CS\_ratios.dat}, \\* {\tt LEP\_2H\_CS\_ratios.dat}, 
{\tt TEV\_1H\_hadCS\_ratios.dat} .
\end{quote}

However, if {\tt whichinput = 'hadr'} and {\tt whichexpt = 'onlyL'}, 
  {\tt HiggsBounds} requires the input files: 

\begin{quote}
{\tt MH\_GammaTot.dat}, {\tt BR\_2H.dat}, {\tt BR\_1H.dat}, 
{\tt LEP\_HZ\_CS\_ratios.dat}, \\* {\tt LEP\_2H\_CS\_ratios.dat} .
\end{quote}

As a third example, if {\tt whichinput = 'hadr'} and {\tt whichexpt = 'onlyT'}, 
 {\tt HiggsBounds} requires the input files: 

\begin{quote}
{\tt MH\_GammaTot.dat}, {\tt BR\_1H.dat}, {\tt TEV\_1H\_hadCS\_ratios.dat} .
\end{quote}

In each of these three examples, {\tt HiggsBounds} 
will also read the file 
\begin{quote}
{\tt additional.dat}
\end{quote}
if it exists.  

As for the subroutine version, if the user does not require processes
involving a particular branching ratio or cross section ratio to be
checked by {\tt HiggsBounds}, that particular array can be filled with
zeros. 
\medskip

\begin{table}
\begin{tabular}{lll}
\hline
{\tt whichinput = 'part'}, &{\tt 'hadr'},  &{\tt 'effC'}\\
\hline
{\tt MH\_GammaTot.dat}   &{\tt MH\_GammaTot.dat} & {\tt MH\_GammaTot.dat}\\ 
{\tt BR\_2H.dat}         &{\tt BR\_2H.dat}       &{\tt effC.dat}  \\
{\tt BR\_1H.dat}         & {\tt BR\_1H.dat}      & {\tt BR\_2H.dat}       \\  
{\tt LEP\_HZ\_CS\_ratios.dat}   & {\tt LEP\_HZ\_CS\_ratios.dat}  
                                                 &  {\tt additional.dat} \\
{\tt LEP\_2H\_CS\_ratios.dat}   &{\tt LEP\_2H\_CS\_ratios.dat} 
                                                 & \\
{\tt TEV\_H\_0jet\_partCS\_ratios.dat} & {\tt TEV\_1H\_hadCS\_ratios.dat} \\   
{\tt TEV\_H\_1jet\_partCS\_ratios.dat} & {\tt additional.dat}  \\  
{\tt TEV\_HW\_partCS\_ratios.dat}     &&\\  
{\tt TEV\_HZ\_partCS\_ratios.dat}     &&\\  
{\tt TEV\_H\_vbf\_hadCS\_ratios.dat} && \\
{\tt additional.dat}               && \\
\hline
\end{tabular}\\
\caption{\sl\label{instructions5}The list of possible input files for each
  value of {\tt whichinput}. Note that some input files may not be
  relevant, depending on the value of {\tt whichexpt}. In this case,
  they are not required. See Table \ref{instructions6} for more
  details. 
\vspace*{5mm}} 
\end{table}

\begin{table}
\begin{tabular}{lllll}
\hline
name of input file & \multicolumn{4}{l}{values of {\tt whichexpt}}     \\
     & \multicolumn{4}{l}{which this file is relevant to}  \\
\hline
                                       &LandT & onlyL& onlyT& singH\\
\hline
{\tt MH\_GammaTot.dat}                 &y     & y     & y     & y     \\
{\tt effC.dat}                         &y     & y     & y     & y     \\    
{\tt LEP\_HZ\_CS\_ratios.dat}          &y     & y     &       & y     \\  
{\tt LEP\_2H\_CS\_ratios.dat}          &y     & y     &       &       \\
{\tt TEV\_H\_0jet\_partCS\_ratios.dat} &y     &       & y     & y     \\   
{\tt TEV\_H\_1jet\_partCS\_ratios.dat} &y     &       & y     & y     \\  
{\tt TEV\_HW\_partCS\_ratios.dat}      &y     &       & y     & y     \\  
{\tt TEV\_HZ\_partCS\_ratios.dat}      &y     &       & y     & y     \\ 
{\tt TEV\_H\_vbf\_hadCS\_ratios.dat}   &y     &       & y     & y     \\ 
{\tt TEV\_1H\_hadCS\_ratios.dat}       &y     &       & y     & y     \\ 
{\tt BR\_1H.dat}                       &y     & y     & y     & y     \\   
{\tt BR\_2H.dat}                       &y     & y     &       &       \\
{\tt additional.dat}  (optional)       &y     & y     & y     & y    \\
\hline
\end{tabular}\\
\caption{\sl\label{instructions6}List of input files, specifying which
  values of {\tt whichexpt} each input file is relevant to (marked by 'y'). 
\vspace*{3mm}} 
\end{table}

{\it Output file format}\\[.3cm]
When the command line version of {\tt HiggsBounds} is used, the output
is written to the file {\tt
  <prefix>HiggsBounds\_results.dat}. A sample of the output is shown in
Fig.~\ref{sampleoutput}. The key to the process numbering is written
to {\tt <prefix>Key.dat}. 


\begin{figure}[ht]
{\scriptsize
\begin{verbatim}
 # generated with HiggsBounds on 31.10.2008 at 11:18
 # settings: LandT, effC
 #
 # column abbreviations
 #   n          : line id of input
 #   Mh(i)      : Higgs boson masses
 #   HBresult   : scenario allowed flag (1: allowed, 0: excluded, -1: unphysical)
 #   chan       : most sensitive channel (see below). chan=0 if no channel applies
 #   obsratio   : ratio [sig x BR]_model/[sig x BR]_limit (<1: allowed, >1: excluded)
 #   ncomb      : number of Higgs bosons combined in most sensitive channel
 #   additional : optional additional data stored in <prefix>additional.dat (e.g. tan beta)
 #
 # channel numbers used in this file
 #           3 : (ee)->(h3)Z->(b b)Z   (LEP table 14b)
 #           4 : (ee)->(h1)Z->(tau tau)Z   (LEP table 14c)
 #         124 : (pp)->W(h1)->l nu (b b)   (CDF Note 9463)
 #         134 : (pp)->h2->tau tau   (arXiv:0805.2491)
 #         157 : (pp)->h1+... where h1 is SM-like  (arXiv:0804.3423 [hep-ex])
 # (for full list of processes, see Key.dat)
 #
 #cols: n    Mh(1)     Mh(2)     Mh(3)   HBresult  chan    obsratio     ncomb  additional(1)
 #
        1   359.121   271.963   134.929         1   134    0.212206E-03     1    0.246862   
        2   75.0123   92.8677   71.9716         1     4    0.306172E-01     1    0.714964   
        3   136.293   345.483   330.026         1   124    0.640713E-01     1    0.434594   
        4   111.377   220.765   51.7469         1     3    0.162811         1    0.727173   
        5   186.131   355.002   146.448         0   157     15.2354         1    0.230522   
\end{verbatim}
}
\caption{\sl\label{sampleoutput} Sample output file (written to {\tt
    <prefix>HiggsBounds\_Results.dat})} 
\end{figure}

\pagebreak

{\it Examples}\\[.3cm]
The {\tt HiggsBounds} package includes a full set of sample input files for
the case $n_H=3$, contained in the folder {\tt example}. Each filename is
prefixed with {\tt nH3\_}. To run the command-line version of 
{\tt HiggsBounds} with these files as input, use, for example,
\begin{verbatim}
./configure
make
./HiggsBounds LandT effC 3 'example/nH3_'
\end{verbatim}
where the values of {\tt whichexpt} and {\tt whichinput} can be varied
as desired.

\subsection{Website version}

The website allows the user to select the required number of neutral Higgs
bosons and then generates a {\tt html} form accordingly. The values of 
{\tt whichinput} and {\tt whichexpt} can be chosen and the appropriate
theoretical input entered. {\tt HiggsBounds} will then be called with these
settings and the result outputted to screen. The website version contains
the additional feature that it notifies the user about the processes with
the second and third highest statistical sensitivities and the values of
{\tt obsratio} for these processes. This is designed to give guidance to the
user who, for example, wishes to find an excluded region iteratively by
adjusting the input quantities.

The website also contains a selection of pre-filled {\tt html} forms as
examples, including entries for the Standard Model, the fermiophobic Higgs
model and the MSSM with real and complex parameters.

\section{Examples of use}
\label{sec:examples}

\subsection{Re-evaluation of SM exclusion limit with an improved SM
prediction}

Recently, an improved prediction for the Higgs-boson production
via gluon fusion appeared \cite{new-gluon-fusion-prediction}, 
which incorporates previously  
unknown mixed QCD--electroweak radiative corrections to this process
as well as other known relevant corrections
\cite{new-gluon-fusion-prediction-using}.
It turns out that those corrections are relevant, increasing the 
SM cross section prediction by 7--10\% compared to the cross section
prediction currently used by the Tevatron 
collaborations, when the MRST 2006 NNLO parton distribution
functions (PDFs)~\cite{MRST2006} are used. 
However, very recently, an updated set of PDFs, 
MSTW 2008 \cite{MSTW2008}, appeared which show a 
decrease in the low-$x$ gluon distribution.
In an updated version of Ref.~\cite{new-gluon-fusion-prediction} the
combined effect of using the improved prediction for the SM cross section
and the latest set of MSTW 2008 NNLO PDFs has been found to result in a
downward shift of the SM cross section prediction 
in the region where the Tevatron searches have the highest sensitivity
by 4--6\% with respect to the cross section prediction
used by the Tevatron collaborations.

\begin{figure}[t]
\centerline{\includegraphics{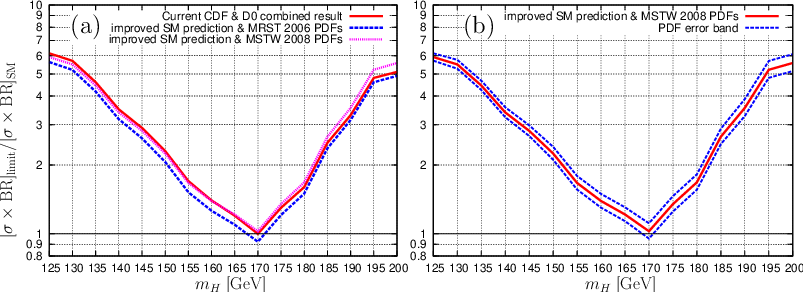}}
    \caption{\sl
	\label{fig:SM-reanalysis} 
	Ratio $Q_\OBS/Q_\SM$ for the most sensitive
	channel as a function of the Higgs mass $m_H$:
	(a) The current best available SM combined analyses of CDF and D\O\
	(solid line) is compared to a re-evaluation using the 
	recently improved SM prediction for the gluon fusion process
	with MRST 2006 PDFs (dashed line) and MSTW 2008 PDFs (dotted line)
	\cite{new-gluon-fusion-prediction};
	(b) The impact of the PDF uncertainty on the exclusion 
	limit using the improved SM prediction
	\cite{new-gluon-fusion-prediction} is shown.
        }
\end{figure}

With {\tt HiggsBounds} the impact of such
changes in the SM cross section prediction on the 
SM exclusion limit can easily be assessed.
Fig.~\ref{fig:SM-reanalysis} shows the result of a
re-evaluation of the Tevatron exclusion limits. 
In Fig.~\ref{fig:SM-reanalysis}(a), 
the limit obtained using the cross section predictions  of
\cite{new-gluon-fusion-prediction} with MRST 2006 NNLO PDFs 
and MSTW 2008 NNLO PDFs
are compared with
the published limits from the 
CDF and D\O\ SM combined analyses 
(\cite{CDF-D0-combined-2} for $m_H < 155\,\gev$ 
and \cite{CDF-D0-combined-3} for $m_H > 155\,\gev$) 
which do not include this improved prediction.
While in the published Tevatron result the cross section 
limit just touches the SM prediction for $m_H\approx 170\,\gev$,
the updated SM cross section prediction exceeds the cross section
limit within an interval of about 5 GeV if MRST 2006 PDFs
are used. 
However, when the most up-to-date MSTW 2008 PDFs are used,
the cross section prediction is lower than the one obtained by
the Tevatron collaborations for $m_H$ above about $160\,\gev$, 
resulting in a reduction of the exclusion power. 
With the updated prediction, no value of the
SM Higgs boson mass is excluded at the 95\% C.L. 
Fig.~\ref{fig:SM-reanalysis}(b) shows how the comparison between the 
SM cross section prediction and the exclusion limits from the Tevatron
is affected by PDF uncertainties. The solid line shows the result based
on the updated SM cross section prediction using MSTW 2008 PDFs, while
the dashed lines indicate the PDF uncertainties as given in 
\citere{new-gluon-fusion-prediction}. It can be clearly seen that currently 
no firm exclusion of any SM Higgs mass values at the 95\% C.L.\
can be established from
the exclusion bounds obtained at the Tevatron once PDF
uncertainties on the SM cross section prediction are taken
into account.

\subsection{Fourth Generation Model}
\label{subsec:FGM}

A very simple example of physics beyond the SM is a
model which extends the SM by a fourth generation of heavy fermions \cite{extra-gen-review}.
In particular, the masses of the 4th generation quarks and leptons
are assumed to be (much) heavier than the mass of the top-quark.
In this case, the effective coupling of the Higgs boson
to two gluons is three times larger than in the SM.
No other coupling, relevant to LEP and Tevatron searches, changes
significantly\footnote{
The effective Higgs-photon-photon coupling is suppressed 
by a factor of about $\frac{1}{9}\Gamma_\TOT^\MOD(H)/\Gamma_\TOT^\SM(H)$
for a light Higgs $H$ \cite{four-gen-and-Higgs}.
The exact suppression factor is irrelevant here, as the Higgs
decay into $\gamma\gamma$ can only lead to relevant search topologies
at the Tevatron if it is enhanced by more than an order of magnitude
compared to the SM.}. 
Essentially, only the partial decay width
$\Gamma(H\to gg)$ changes by a factor of 9 and, with it, the 
total Higgs width and therefore all the decay branching ratios 
\cite{four-gen-and-Higgs}.

\begin{figure}[ht]
\begin{footnotesize}
{\tt
\hspace*{1cm}	program SM\_vs\_fourth\_generation\_model\\
\hspace*{1cm}	implicit none\\
\hspace*{1cm}	integer nH,HBresult,chan,ncombined\\
\hspace*{1cm}	real*8 obsratio,SMGammaTotal,GammaTotal,SMgamma\_hgg,\\
\hspace*{.6cm}     \& \ \ \ 
                      Gamma\_hgg,SMGamma\_h,SMBR\_Hgg,Mh,g2hjbb,g2hjtautau,\\
\hspace*{.6cm}     \& \ \ \ 
                      g2hjWW,g2hjZZ,g2hjgaga,g2hjgg,g2hjhiZ,BR\_hjhihi\\[.1cm]
\hspace*{1cm}	nH=1\\
\hspace*{1cm}	call initialize\_HiggsBounds(nH,'LandT')\\
\hspace*{1cm}	open(10,file='example-SM-results.dat')\\
\hspace*{1cm}	open(20,file='example-4thGen-results.dat')\\[.1cm]
\hspace*{1cm}	do Mh=90,200,5\\
\hspace*{1cm}	\  SMGammaTotal=SMGamma\_h(Mh)\\
\hspace*{1cm}	\  SMgamma\_hgg=SMBR\_Hgg(Mh)*SMGammaTotal\\
\hspace*{1cm}	\  Gamma\_hgg=9d0*SMBR\_Hgg(Mh)*SMGammaTotal\\
\hspace*{1cm}	\  GammaTotal=SMGammaTotal-SMgamma\_hgg+Gamma\_hgg\\[.1cm]
\hspace*{1cm}	\  g2hjbb=1d0\\
\hspace*{1cm}	\  g2hjtautau=1d0\\
\hspace*{1cm}	\  g2hjWW=1d0\\
\hspace*{1cm}	\  g2hjZZ=1d0\\
\hspace*{1cm}	\  g2hjgaga=1d0\\
\hspace*{1cm}	\  g2hjgg=1d0\\
\hspace*{1cm}	\  g2hjhiZ=0d0\\
\hspace*{1cm}   \  BR\_hjhihi=0d0\\[.1cm]
\hspace*{1cm}	\  call run\_HiggsBounds\_effC(nH,Mh,SMGammaTotal,\\
\hspace*{.6cm}     \& \ \ \ 
                   g2hjbb,g2hjtautau,g2hjWW,g2hjZZ,g2hjgaga,g2hjgg,g2hjhiZ,\\
\hspace*{.6cm}     \& \ \ \ BR\_hjhihi,HBresult,chan,obsratio,ncombined)\\
\hspace*{1cm}	\  write(10,*) Mh,HBresult,chan,obsratio\\[.1cm]
\hspace*{1cm}	\  g2hjgaga=1d0/9d0*GammaTotal/SMGammaTotal\\
\hspace*{1cm}	\  g2hjgg=9d0\\[.1cm]
\hspace*{1cm}	\  call run\_HiggsBounds\_effC(nH,Mh,GammaTotal,\\
\hspace*{.6cm}     \& \ \ \ 
                   g2hjbb,g2hjtautau,g2hjWW,g2hjZZ,g2hjgaga,g2hjgg,g2hjhiZ,\\
\hspace*{.6cm}     \& \ \ \ BR\_hjhihi,HBresult,chan,obsratio,ncombined)\\
\hspace*{1cm}	\  write(20,*) Mh,HBresult,chan,obsratio\\
\hspace*{1cm}	enddo\\[.1cm]
\hspace*{1cm}	close(10)\\
\hspace*{1cm}   close(20)\\
\hspace*{1cm}   call finish\_HiggsBounds\\[.1cm]
\hspace*{1cm}	end
}
\end{footnotesize}
\caption{\sl\label{fig:fourth-gen-code} Fortran77 code 
{\tt example-SM\_vs\_4thGen.F}  which 
calculates the ratio $Q_\MOD/Q_\OBS$ ({\tt obsratio}) for the most sensitive
Higgs search topology for the SM and a simple Fourth Generation Model
as a function of the Higgs mass ({\tt Mh}). This program has been used to
produce the plot shown in Fig.~\ref{fig:4th-gen-model}.
}
\end{figure}

\begin{figure}[t]
\centerline{\includegraphics{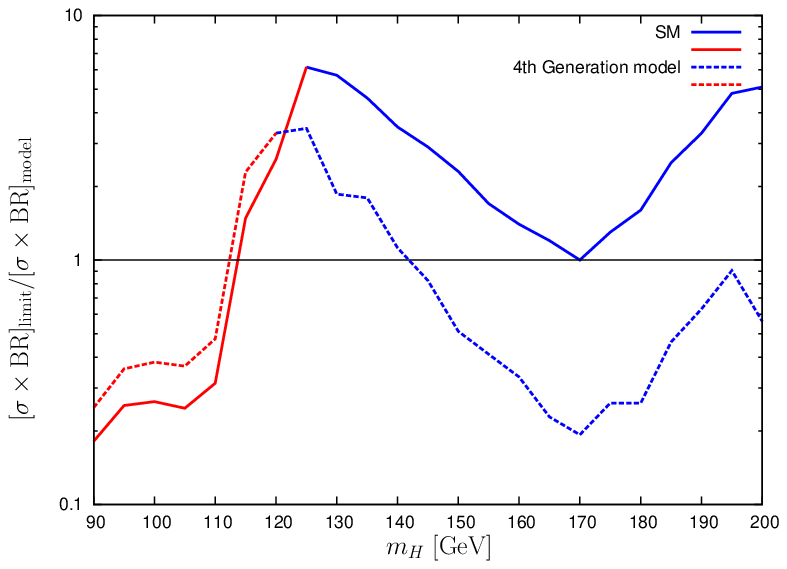}}
    \caption{\sl
	\label{fig:4th-gen-model} 
	Ratio $Q_\OBS/Q_\MOD$ for the most sensitive
	channel as a function of the Higgs mass $m_H$:
	4th Generation Model versus SM. 
	The colours indicate whether the most sensitive
	search channel is from 
	\GNUPlotA{LEP} (lighter grey) 
        or the \GNUPlotC{Tevatron} (darker grey).
        }
\end{figure}

We can easily calculate the new total decay width and the 
relevant decay branching ratios using the above information
and the built-in functions to calculate SM quantities 
using the relations:
\begin{align*}
\Gamma_\SM(H\to gg) & = \BR_\SM(H\to gg)\:\Gamma_\TOT^\SM(H)\,,\\ 
\Gamma_\MOD(H\to gg) &= 9\:\Gamma_\SM(H\to gg)\,,\\
\Gamma_\TOT^\MOD(H) &= \Gamma_\TOT^\SM(H) - \Gamma_\SM(H\to gg)
	+ \Gamma_\MOD(H\to gg)\,.
\end{align*}
It is then very simple to test the model with {\tt HiggsBounds}
using the effective coupling input.

In Fig.~\ref{fig:fourth-gen-code} we show a Fortran 77 code example
which uses {\tt HiggsBounds} for the calculation of the
Higgs mass dependence of the ratio $Q_\MOD/Q_\OBS$
for the SM and our simple Fourth Generation Model scenario.
From the results returned by the subroutine {\tt run\_HiggsBounds\_effC}, 
the set ({\tt mh}, {\tt HBresult}, {\tt chan}, {\tt obsratio})
is stored in a file, separately for the two models.

The results obtained by this sample program are 
presented in Fig.~\ref{fig:4th-gen-model}, which shows
the ratio $(Q_\MOD/Q_\OBS)^{-1} =
[\sigma\times\text{BR}]_{\text{obs. limit}}/[\sigma\times\text{BR}]_{\text{model}}$
as a function of the Higgs mass $m_H$ for the SM (solid lines) and the 
Fourth Generation Model (dashed lines).
For the SM cross section we use here the prediction employed by the
Tevatron collaborations, see the discussion above concerning an update
of this prediction.
Exclusion at 95\% C.L.
is given if the value of $(Q_\MOD/Q_\OBS)^{-1}$ drops below 1.
Both curves have sections in red colour (lighter grey), 
where the LEP results 
provide the most sensitive search topology, and in blue colour (darker
grey), 
where the most sensitive search topology is provided by 
the Tevatron results.
As expected, the LEP results have the highest sensitivity
in the low-mass region
up to the kinematical limit of about 115~GeV, while 
for higher mass values
the Tevatron results take over.

The SM results show the familiar exclusion of Higgs masses below 
$\approx 115 \gev$ from LEP and the small excluded region
at about $m_H = 170 \gev$ from the latest Tevatron SM combined
analysis \cite{CDF-D0-combined-3} using the results of the slightly more 
conservative results obtained by the Bayesian method (see the discussion
above for an update of the cross-section prediction employed here).
For the Fourth Generation Model the LEP exclusion is a bit 
weaker than for the SM, because at low Higgs mass 
the larger branching ratio for Higgs decays into gluon pairs 
diminishes the $b\bar b$ and $\tau^+\tau^-$ branching ratios.
In the region above 115 GeV, the most sensitive search topology
for the Fourth Generation Model Higgs 
is $p\bar p\to H \to W^+W^-\to l^+ l'^-$ throughout the 
displayed range. Because of the enhancement of the 
signal cross section due to the enhanced coupling of the Higgs boson
to gluons, the region excluded by the Tevatron results extends
from about 145 GeV to the end of the displayed range.
Because of the SM-likeness criterion,
the single topology analyses \cite{D0-5757,CDF-0809-3930}
apply, instead of the SM combined analysis, for 
the Fourth Generation Model. 
Disregarding this criterion would change the $(Q_\MOD/Q_\OBS)^{-1}$ values
in the Tevatron exclusion region to almost exactly 
1/9 times the SM values.

\subsection{MSSM}

{\em Theoretical background}

\noindent
The MSSM 
Higgs sector consists of two scalar weak isospin doublets and
possesses therefore
five physical Higgs bosons. At tree-level these are
the $\cp$-even $h$ and $H$ (with $\mh < \mH$), the $\cp$-odd $A$ and the
charged $H^\pm$. 
The Higgs sector of the MSSM can be expressed at lowest
order in terms of $\MZ$, $\MA$ (or $\MHp$) and 
$\tb \equiv v_2/v_1$, 
the ratio of the vacuum expectation values
of the two doublets.
All other masses and
mixing angles can therefore be predicted.
The tree-level masses and couplings are strongly
affected by higher-order corrections~\cite{MSSMHiggsRev}. 
In the real MSSM (rMSSM), i.e.\ the model without $\cp$-violation, 
the higher-order corrected masses are denoted as $\Mh$, $\MH$
and $\MA$. In the MSSM with complex phases (cMSSM) the three neutral
Higgs bosons can mix and the masses are denoted as 
$\Mhe$, $\Mhz$, $\Mhd$ (with $\Mhe < \Mhz < \Mhd$).

In order to ensure that the external Higgs bosons have the correct on-shell
properties, the tree-level Higgs states have to be supplemented with
Higgs-propagator corrections (which are often the dominant corrections).
This is done with the help of the 
$\matr{Z}$~matrix~\cite{mhcMSSMlong,Hahn:2006np},
a non-unitary matrix,
which mixes between the Higgs tree-level mass eigenstates (in the order
$h,H,A$) and the loop-corrected mass eigenstates 
\begin{align}
\begin{pmatrix} \He \\ \Hz \\ \Hd \end{pmatrix} = \matr{Z}\cdot
\begin{pmatrix} h \\ H \\ A \end{pmatrix} . 
\end{align} 
In the rMSSM, setting
$\He = h$, $\Hz = H$, $\Hd = A$, one has 
\begin{align} Z_{31} = Z_{32} = Z_{13} =
Z_{23} = 0, \; Z_{33} = 1 . \label{ZrMSSM} 
\end{align} 
In the following we
give the formulas valid for the more general cMSSM, where the special
case of the rMSSM can
be derived with the help of \refeq{ZrMSSM}. The normalised cross sections
$\sigma^{\rm norm}_{h_iZ}$, $\sigma^{\rm norm}_{h_jh_i}$,
defined through Eqs.~(\ref{basic input}),
(\ref{eeHZ-ref}) and (\ref{eeHH-ref}) as part of the input
needed by {\tt HiggsBounds},
can then be found using
the elements of the matrix $\matr{Z}$ 
\BEA 
\sigma^{\rm norm}_{h_iZ}&=&\left|Z_{i1} \sin (\beta-\alpha)+Z_{i2} \cos
  (\beta-\alpha)\right|^2\nonumber\\ 
\sigma^{\rm norm}_{h_jh_i}&=& \left|Z_{i 3} ( Z_{j 1} \cos 
  (\beta-\alpha) - Z_{j 2} \sin (\beta-\alpha) )\right.\\ 
 &&\left.- Z_{j 3} ( Z_{i 1} \cos (\beta-\alpha) -Z_{i 2} \sin 
  (\beta-\alpha) )\right|^2 ,
\label{coupZ}
\EEA
where $\al$ is the angle that diagonalises the $\cp$-even Higgs sector at
tree-level. 
It should be noted that for two identical Higgs bosons this leads to 
$\sigma^{\rm norm}_{h_ih_i}=0$ and that the following approximate relations hold:
\BEA
\sigma^{\rm norm}_{h_1Z}+\sigma^{\rm norm}_{h_2Z}+\sigma^{\rm norm}_{h_3Z}&\simeq&1 ,\\
\sigma^{\rm norm}_{h_kZ}&\simeq&\sigma^{\rm norm}_{h_jh_i},
\label{eq:MSSMinequal}
\EEA 
where $i,j,k$ are all different\footnote{Note that these relations become
equalities if a unitary Higgs mixing matrix is used. See, e.g.,
\citere{mhcMSSMlong}.}.
\medskip

{\em Limits for the real MSSM }\\[.3cm]
In \citere{LEP-MSSM-Higgs-analysis} the final results for the LEP Higgs
searches in the MSSM were published. In the case of the rMSSM they had
been obtained in four benchmark scenarios: the \mhmax, no-mixing,
small~$\aeff$ and gluophobic Higgs scenarios (including also some
variations), see \citere{benchmark2} for 
a motivation and detailed definition of these scenarios. The computer
code used at that time was {\tt FeynHiggs\,2.0}, and in most cases the
top-quark mass was fixed to $\mt = 174.3 \gev$. The limits had been obtained
by a combination of all available channels. 

Here we present the limits obtained by {\tt HiggsBounds} using the
latest version (2.6.4) of {\tt FeynHiggs}
to calculate the required model input and choosing different values
for $\mt$.
Note that the limits given here are in the narrow-width approximation,
as discussed in Section \ref{subsec:general approach}.

\begin{figure}[t]
\centerline{\includegraphics{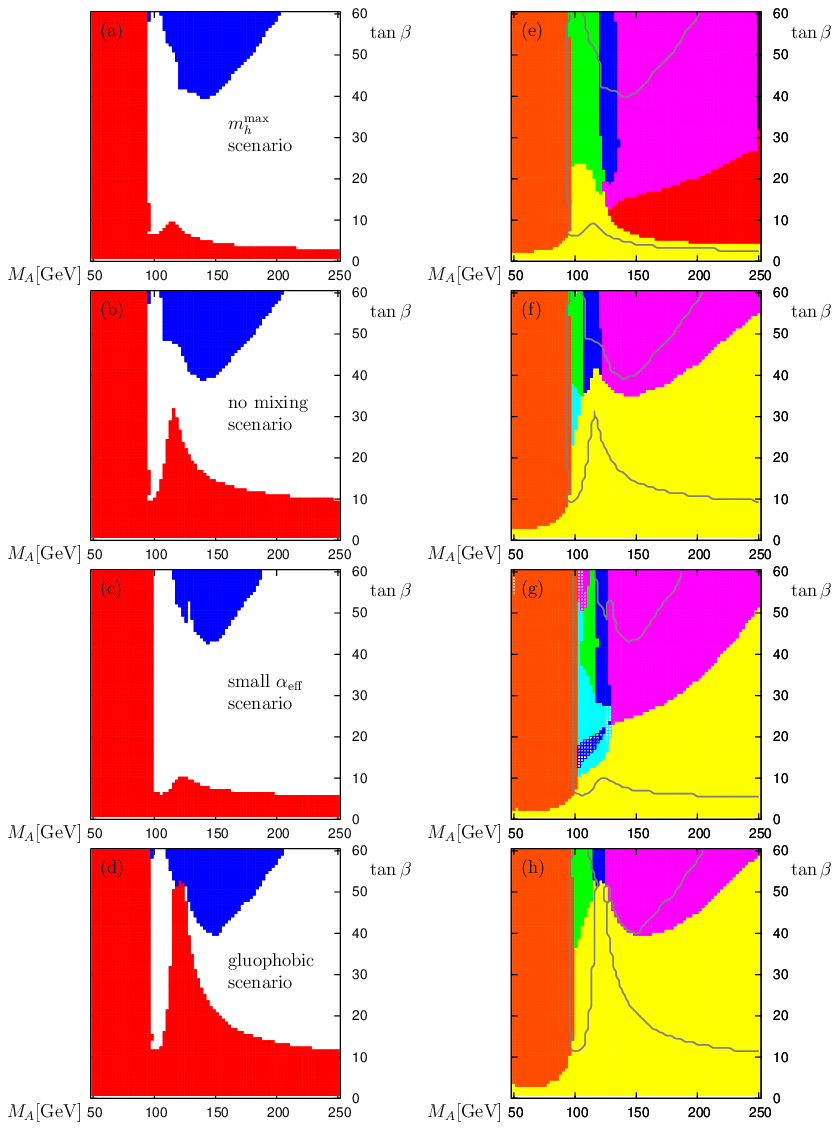}}
    \caption{\sl
	\label{fig:rMSSM1} 
	Left column [(a),(b),(c),(d)]: 95\% C.L. level excluded
        regions in the
	$\MA$--$\tb$ plane for the the four rMSSM benchmark scenarios.
	The colours indicate exclusion from LEP (\GNUPlotA{$\blacksquare$}) 
	or the Tevatron (\GNUPlotC{$\blacksquare$}).
	Right column [(e),(f),(g),(h)]: channel of highest 
	statistical sensitivity:
	yellow (\GNUPlotF{$\blacksquare$}): 
		$e^+ e^- \to hZ, h\to b\bar b$;
	cyan (\GNUPlotE{$\blacksquare$}): 
		$e^+ e^- \to HZ, H\to b\bar b$;
	cyan (\GNUPlotE{$\boxdot$}): 
		$e^+ e^- \to hZ/HZ, h/H\to b\bar b$;
	orange (\GNUPlotH{$\blacksquare$}/\GNUPlotH{$\boxdot$}): 
		$e^+ e^- \to hA \to b\bar b b\bar b/ b\bar b \tau^+\tau^-$;
	green (\GNUPlotB{$\blacksquare$}): 
		$p \bar p \to h/A \to \tau^+\tau^-$ \cite{CDF-9071};
	blue (\GNUPlotC{$\blacksquare$}): 
		$p \bar p \to h/H/A \to \tau^+\tau^-$ \cite{CDF-9071};
	magenta (\GNUPlotD{$\blacksquare$}): 
		$p \bar p \to H/A \to \tau^+\tau^-$ \cite{CDF-9071};
	magenta (\GNUPlotD{$\boxdot$}): 
		$p \bar p \to A \to \tau^+\tau^-$ \cite{CDF-9071};
	red (\GNUPlotA{$\blacksquare$}): 
		$p \bar p \to h W\to b \bar b l\nu$ \cite{CDF-9463};
	black (\GNUPlotG{$\blacksquare$}): 
		$p \bar p \to H/A \to \tau^+\tau^-$ \cite{D0-0805-2491};
	blue (\GNUPlotC{$\boxdot$}): 
		$p \bar p \to H \to W^+ W^-$ \cite{D0-5757}.
        The contour borders the excluded regions. 
        }
\vspace{-3em}
\end{figure}

In the left column of \reffi{fig:rMSSM1} we show the LEP and Tevatron
exclusion bounds as obtained 
with {\tt HiggsBounds} in the \mhmax (a), no-mixing (b),
small~$\aeff$ (c) and gluophobic Higgs scenario (d)
in the $\MA$--$\tb$ plane for the
current top-quark mass value, $\mt = 172.4 \gev$~\cite{mt1724}.
In the right column of 
\reffi{fig:rMSSM1} we show the most sensitive channels for each point
in the $\MA$--$\tb$ plane for the same four scenarios~(e)--(h)
(the colour code is given in the caption).

The four scenarios show similar features in the excluded regions, as
presented in the left-hand column. Very
low $\MA$ values are excluded by LEP searches for all $\tb$ values.
This is due to the channel $e^+e^- \to hA \to b \bar b \, b \bar b$ as
can be seen in the right column of \reffi{fig:rMSSM1}. 
In this region the $hZZ$ coupling is suppressed. Only at very low $\tb$
values the channel $e^+e^- \to hZ, h \to b \bar b$ has the highest statistical sensitivity, but still with a somewhat suppressed $hZZ$ coupling.
At around 
$\MA \approx 100 \gev$ the kinematical limit for the $hA$ channel is
reached and other channels take over. There the exclusion drops to a
relatively low excluded $\tb$ value, slightly different in the four
scenarios. 
At around $\MA \approx 115 \gev$ the exclusion in the 
$e^+e^- \to hZ, h \to b \bar b$ channel becomes SM-like, and Higgs
bosons up to the kinematical limit can be excluded by LEP searches,
reaching $\tb = 8, 30, 10, 50$ in the \mhmax, no-mixing, small~$\aeff$
and gluophobic Higgs scenario, respectively. In the latter one also
larger values could be excluded, but the Tevatron channels become more
important, see below.
For larger $\MA$ values the LEP exclusion drops to a lower $\tb$ value
between 3~and~11, depending on the scenario. These $\tb$ values
correspond to $\Mh \approx 208 \gev - \MZ$, i.e.\ the kinematical limit
and are excluded for all $\MA$ values above $\sim 115 \gev$.

The Tevatron searches exclude a parameter space, depending on the
scenario, between $\MA = 110 \gev$ 
and $170 \gev$, and between $\tb = 39$ and 60~(where we stopped our
scan, but in principle larger values can be excluded). The relevant
channels are $p \bar p \to h/H/A + X, h/H/A \to \tau^+\tau^-$, see
the right-hand column of \reffi{fig:rMSSM1}.
The variation in the colour here only indicates which Higgs bosons
contribute together to the signal. At low $\MA$ (green) one has
$\Mh \approx \MA$, at intermediate $\MA$ (dark blue) we find 
$\Mh \approx \MA \approx \MH$, and at higher $\MA$ values one
generically has $\MA \approx \MH$. Consequently, two or three
neutral Higgs bosons can contribute to the signal.
Two other Tevatron channels have the largest exclusion potential
in some part of the $\MA$--$\tb$ parameter space. 
Within the \mhmax\ scenario $p \bar p \to h W\to b \bar b l\nu$ is the most
sensitive channel for $\MA \gsim 130 \gev$ and intermediate $\tb$, see
plot (e) in \reffi{fig:rMSSM1}. 
In the small~$\aeff$ scenario $p \bar p \to H \to W^+ W^-$ 
with the $W$s decaying leptonically is the most sensitive
channel for low $\MA$ and
intermediate $\tb$, see plot (g) of \reffi{fig:rMSSM1}. 

\begin{figure}[t]
\centerline{\includegraphics{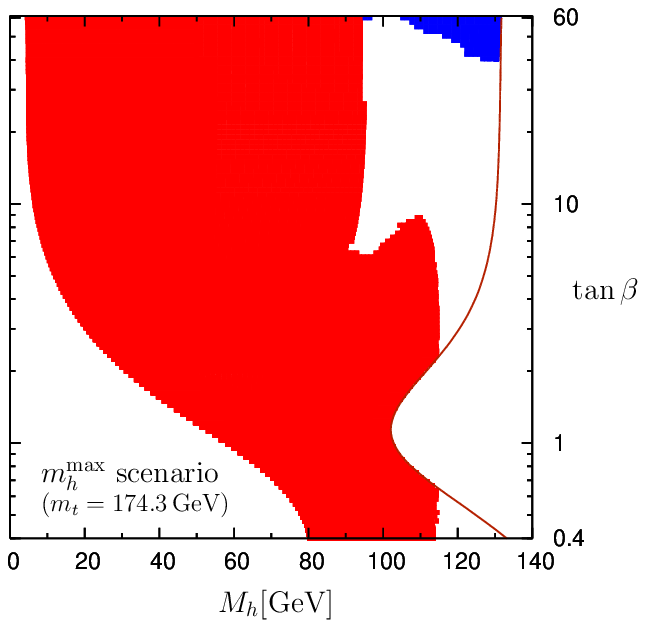}}
    \caption{\sl
	\label{fig:rMSSM-mh0-tb-oldMT} 
	Parameter region in the $\Mh$--$\tb$ plane
	for the $m_h^\MAX$ scenario of the rMSSM
	which is excluded at the 95\% C.L. 
	The mass of the top quark has been set to 174.3 GeV.
	The colours indicate whether the most sensitive
	search channel which leads to exclusion is from 
	\GNUPlotA{LEP} (lighter grey) 
        or the \GNUPlotC{Tevatron} (darker grey).
	The solid line shows the theoretical upper bound 
	on $\Mh$ in the $m_h^\MAX$ scenario.
        }
\end{figure}

As mentioned before, the original exclusion bounds in
\citere{LEP-MSSM-Higgs-analysis} had been obtained for 
$\mt = 174.3 \gev$. The most prominent effect of increasing the
top-quark mass from $\mt = 172.4 \gev$ to $\mt = 174.3 \gev$
is an increase in $\Mh$ with 
$\De\Mh/\De\mt \lsim 1$~\cite{tbexcl}. This reduces the reach in the
$\MA$--$\tb$ plane at LEP for the $e^+e^- \to hZ$ channels. This effect
can be observed to some extent in all four scenarios. 
In \reffi{fig:rMSSM-mh0-tb-oldMT} we show the exclusion bounds as
obtained with {\tt HiggsBounds} with $\mt = 174.3 \gev$ in the
\mhmax\ scenario in the $\Mh$--$\tb$ plane. 
This plot can be compared with Fig.~9(b) in the
original LEP publication~\cite{LEP-MSSM-Higgs-analysis}, which, however,
terminated at $\tb = 40$, whereas the plot shown here extends up to $\tb = 60$.
The values of $\Mh$ reached in \reffi{fig:rMSSM-mh0-tb-oldMT} are
smaller by $\sim 2 \gev$ as compared to the original LEP figure, which
is due to the newer version of {\tt FeynHiggs} employed here.
Apart from this, the LEP exclusion areas are almost identical in the two
figures. This shows how well in this case our procedure (based on the
LEP data for single topological cross sections) 
reproduces the full LEP analysis where different search channels were
combined. 
In addition, we also find a 
region excluded by the Tevatron Higgs searches at $\tb \gsim 40$ and 
$\Mh \gsim 100 \gev$. These results from the Tevatron were
not yet available when 
\citere{LEP-MSSM-Higgs-analysis} was published. 
\medskip

{\em LEP limits for the complex MSSM in the CPX scenario}\\[.3cm]
\citere{LEP-MSSM-Higgs-analysis} also published the results of the LEP Higgs
searches in the CPX scenario, which is a $\cp$-violating MSSM benchmark
scenario~\cite{Carena:2000ks}. This analysis incorporated Higgs sector
predictions from the programs {\tt CPH}~\cite{Carena:2000ks,Carena:2000yi}
and {\tt FeynHiggs} 2.0~\cite{feynhiggs,mhiggslong,mhiggsAEC,mhcMSSMlong}.

More recently, in \citere{Williams:2007dc}, a preliminary version of 
{\tt HiggsBounds} was
used in conjunction with full 1-loop vertex corrections for the $h_j \to h_i
h_i$ and $h_j \to f \bar{f}$ decays to examine the effect of these
corrections on the LEP exclusions in the CPX scenario. In
\citere{Williams:2007dc}, Higgs masses and the Higgs propagator corrections
(as contained in the $\matr{Z}$~matrix) were calculated using renormalised
Higgs self-energies from {\tt FeynHiggs} 2.6.4. 

\begin{figure}[tb]
\begin{tabular}{cc}
\includegraphics[width=.48\linewidth,angle=0]{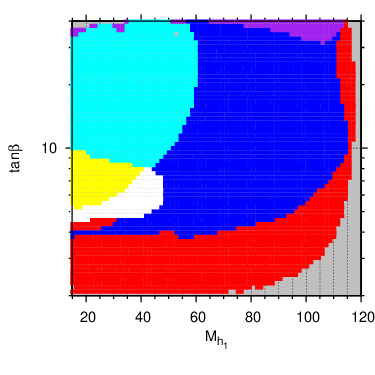}&
\includegraphics[width=.48\linewidth,angle=0]{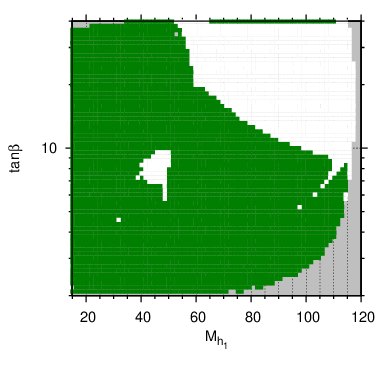}
\\[-.5em]
(a)&(b)
\end{tabular}
\caption{\sl Coverage of the LEP Higgs searches in the $M_{H_1}$--$\tb$ plane
of the CPX scenario. Plot (a) shows the channels that are
predicted to have the highest statistical sensitivity for setting an
exclusion limit.
The colour codings are: 
red $(\GNUPlotA{\blacksquare})=$ $(h_1 Z)\to(b \bar b Z)$,
blue $(\GNUPlotC{\blacksquare})=$ $(h_2  Z)\to(b \bar b Z)$,
white $(\square)=$ $(h_2  Z)\to(h_1 h_1 Z)\to(b \bar b b \bar b Z)$,
cyan $(\GNUPlotE{\blacksquare})=$ $(h_2 h_1)\to(b \bar b b \bar b)$,
yellow $(\GNUPlotF{\blacksquare})=$ $(h_2 h_1)\to(h_1 h_1 h_1)\to(b \bar b b \bar b b \bar b)$,
purple $(\Purple{\blacksquare})=$ other channels.
Plot (b) shows the parameter regions excluded
at the 95\% C.L.\
by the topological cross section limits obtained at
LEP~\cite{LEP-SM-Higgs-analysis, LEP-MSSM-Higgs-analysis}.
The colour codings are:
green $(\KarinaGreen{\blacksquare})=$ excluded, white $(\square)=$ unexcluded.
In both plots, the grey $(\KarinaGrey{\blacksquare})$ peripheral areas 
are theoretically inaccessible 
using the input parameters associated with the CPX scenario.
\label{fig:CPX}
}
\end{figure}

The results shown in \reffi{fig:CPX} have been calculated using the
conventions\footnote{Note that \citere{LEP-MSSM-Higgs-analysis} and
\citere{Williams:2007dc} use slightly different definitions of the CPX
scenario (see \citere{Williams:2007dc} for a full description of these
differences).} and method described in \citere{Williams:2007dc}. However,
the results shown here include two significant updates: we use $m_t=172.6
\gev$ rather than $m_t=170.9 \gev$ and we use a newer version of 
{\tt FeynHiggs} to calculate the renormalised Higgs self-energies, which
contains the correct combination of the 
$\tan\beta$-enhanced contributions to the Higgs-boson self-energies in 
the complex
MSSM with the one-loop corrections for all options concerning the two-loop
contributions.

In \reffi{fig:CPX} (a), we show the channels with the highest statistical
sensitivity for each point in CPX parameter space. The process 
$(h_1 Z)\to(b\bar b Z)$ (red)\footnote{All processes discussed 
in this subsection are LEP
processes. Therefore, for brevity, we do not include the initial state
explicitly in the process descriptions.} dominates in regions where the
$h_1ZZ$ effective coupling is significant. Adjacent to this region is an
area of parameter space with a relatively large $ h_2ZZ$ effective coupling
and therefore the processes $(h_2 Z)\to(b \bar b Z)$ (blue) and $(h_2
Z)\to(h_1 h_1 Z)\to(b \bar b b \bar b Z)$(white) have the highest
statistical sensitivity in this region. The processes $(h_2 h_1)\to(b \bar b
b \bar b)$(cyan) and $(h_2 h_1)\to(h_1 h_1 h_1)\to(b \bar b b \bar b b \bar
b)$ (yellow) dominate in the region where the $h_3ZZ$ effective coupling is
relatively large
 (and therefore also the $h_1h_2Z$ effective coupling, as predicted
by \refeq{eq:MSSMinequal}). 

In the region of \reffi{fig:CPX} (a) where $M_{h_1}\lsim 45 \gev$, $\tb\gsim
4$, the total decay width of $h_2$ is dominated by the partial decay width
$h_2 \to h_1 h_1$. This has a very significant effect on the exclusion
prospects in this region. As \reffi{fig:CPX} (a) shows, this region is
partly covered by the $(h_2 Z)\to(h_1 h_1 Z)\to(b \bar b b \bar b Z)$
(white) and $(h_2 h_1)\to(h_1 h_1 h_1)\to(b \bar b b \bar b b \bar b)$
(yellow) processes, which directly involve the $h_2\to h_1 h_1$ decay.
However, searching for these topologies is very challenging experimentally
as, for example, the 
$(h_2 h_1)\to(h_1 h_1 h_1)\to(b \bar b b \bar b b \bar b)$ process 
gives rise to complex final states involving six jets.
Additionally, processes directly involving the $h_2 \to b \bar{b}$ decay
(such as $(h_2 Z)\to(b \bar b Z)$ (blue)) are more difficult to detect in
this region because of the suppression of the $h_2 \to b \bar{b}$ branching
ratio due to the large $h_2\to h_1 h_1$ partial decay width.

\reffi{fig:CPX} (b) shows the areas of CPX parameter space which have been
excluded by the LEP Higgs searches at the 95 \% C.L. (green). There are two
sizable unexcluded regions\footnote{Note that we do not consider unexcluded
regions to be significant 
if they only occur in a very small Higgs mass range and are situated 
on the borders
of regions where different channels have 
the highest statistical sensitivity. As discussed in Section 2.1, 
we expect HiggsBounds to be less sensitive at
these borders than a dedicated experimental analysis.
} (white), at $M_{h_1}\gsim 50 \gev$, $ \tb \gsim 8
$ and a ``hole'' at $M_{h_1}\approx 45 \gev$, $ \tb \approx 8$. Both of these regions
occur in parts of parameter space where processes involving the topology
$(h_i Z)\to(b\bar b Z)$ predominantly have the highest statistical
sensitivity, and the second of these regions occurs in the part of the CPX
parameter space which, as discussed above, was expected to prove
challenging experimentally due to a suppressed $h_2 \to b \bar{b}$ branching
ratio. 
In addition, the
two unexcluded regions are very sensitive to the slight excess
in the LEP data for 
this topology at $89.6 \gev
\lsim M_{h_i} \lsim 107 \gev$, where the observed limit is more than 1 sigma
above the predicted limit based on simulations with no
signal, see \citere{LEP-MSSM-Higgs-analysis}.

Comparison of \reffi{fig:CPX} (a) with \reffi{fig:CPX} (b) also shows that
there are some unexcluded parameter points along boundaries between areas
where different processes have the highest statistical sensitivity, for
example, at $M_{h_1}\approx 100 \gev$, $ \tb \approx 5$. Since in our
analysis only the single channel having the highest statistical
sensitivity can be used to set the
95\% C.L.\ limit, the statistical information from all other decay and
production channels in this region does not contribute.
In these narrow
regions, an analysis which can utilise more than one experimentally measured
cross section limit for each parameter point (such as that performed in
\citere{LEP-MSSM-Higgs-analysis}) will have a higher exclusion power, and
thus possibly be able to reduce the 
size of the unexcluded region.

Using {\tt HiggsBounds}, it has therefore been possible to investigate the
effects of recent improvements in the theoretical Higgs sector predictions
on the LEP exclusions in the CPX scenario. In particular, since it provides
information about the process with the highest statistical sensitivity, 
{\tt HiggsBounds} should be helpful for the user to obtain some guidance 
on which effects are important for the exclusion or non-exclusion
of certain parts of parameter space.
{\tt HiggsBounds}
also provides a convenient way to investigate the dependence of the
calculation on the model input parameters, such as the effect of varying the
top-quark mass.

\section{Summary}
\label{sec:summary}

We have presented the code {\tt HiggsBounds}, which is a tool to test
whether the predictions arising from the Higgs sector of arbitrary types
of theoretical models are in accordance with the existing exclusion
bounds from the Higgs searches at LEP and the Tevatron. To this end, the 
experimental bounds on topological cross sections made available by the 
LEP and Tevatron collaborations have been incorporated as look-up tables
into the program. In order to utilise this information to set exclusion
bounds on the parameter space of a certain model one first needs to
determine the search channel with the highest statistical sensitivity
for setting an exclusion limit. This is done by comparing the expected limits 
(which have also been provided by the experimental collaborations),
i.e.\ the exclusion bounds that one would obtain in the hypothetical
case of an observed distribution with a pure background shape, with the
model predictions. In order to ensure the correct statistical
interpretation of the obtained exclusion bound as a 95\% C.L., the
analysis has to be restricted to the single channel that possesses the 
highest statistical sensitivity for setting an exclusion limit.
For this channel the program determines whether or not a certain
parameter point is excluded at the 95\% C.L. 

The experimental information incorporated into {\tt HiggsBounds}
consists primarily of limits on single topological cross sections, 
either from a single collaboration or combinations of the results 
from different collaborations. 
Some experimental combinations have been done by the
collaborations only for the special case of a SM-type Higgs boson.
These have the advantage of an enhanced statistical power.
In order to
determine whether experimental results on the searches for the SM Higgs 
can be used for constraining a certain parameter region of a 
new physics scenario, the ``SM
likeness'' of the Higgs boson(s) of the model under study is evaluated.
Most experimental results that are currently available have been
obtained under the assumption that a narrow width approximation is valid
for the Higgs boson under study. 
If this assumption is not valid
the experimental exclusion bounds are modified. The current version of 
{\tt HiggsBounds} is based on exclusion limits obtained in the
narrow-width approximation. The proper treatment of experimental results,
taking into account a non-negligible Higgs-boson width, will be 
incorporated in a forthcoming version of {\tt HiggsBounds}.
The exclusion bounds included in the {\tt HiggsBounds} package will be
updated as new results from the Higgs searches become available.

The predictions of the desired model, needed as input for {\tt HiggsBounds},
have
to be provided by the user. This can be done at various levels
of sophistication, ranging from effective couplings to complete
information on cross sections and branching ratios.
Links to the widely used programs {\tt FeynHiggs} and {\tt CPsuperH} for
Higgs-sector predictions in the MSSM are provided by default.

We have discussed various examples for running {\tt HiggsBounds} and
have given sample files for its input and output. In particular, we have 
considered the cases of the Standard Model, a model with a fourth 
generation of
quarks and leptons and for the MSSM with and without $\cp$-violation.

The {\tt HiggsBounds} code exists both in a 
Fortran 77 and Fortran 90 version. It can be operated in a command line
mode, as a subroutine that can easily be linked to other codes, and
as an online version, see \\
{\tt www.ippp.dur.ac.uk/HiggsBounds}.

\section*{Acknowledgements}
We thank the LEP collaborations ALEPH, DELPHI, L3 and OPAL and the 
Tevatron collaborations CDF and D\O\ for providing us with detailed 
information about their Higgs search results. 
In particular, we are
grateful to Alexander Read and Peter Igo-Kemenes (LEP Higgs Working Group),
Mark~Owen (D\O), Tom~Junk and Matthew~Herndon (CDF)
for very valuable discussions on their experiment's
results.
We thank Terrance Figy for providing us with SM predictions
for the Higgs production cross section via VBF at the Tevatron
separated in $W$- and $Z$-fusion contribution.
We would like to thank Philip Roffe and David Ambrose-Griffith for
valuable computing support, and the IPPP (Durham, UK) 
for hosting the web version of {\tt HiggsBounds}. 
Thanks also go to Lydia Heck, Sophy Palmer and Alison Fowler 
for their assistance in testing the code and to Valery Khoze for
interesting discussions. 
KW was supported by the Helmholtz Alliance HA-101 
'Physics at the Terascale'.
This work has been supported 
in part by the European Community's Marie-Curie Research
Training Network under contract MRTN-CT-2006-035505
`Tools and Precision Calculations for Physics Discoveries at Colliders'
(HEPTOOLS) and MRTN-CT-2006-035657
`Understanding the Electroweak Symmetry
Breaking and the Origin of Mass using the First Data of ATLAS'
(ARTEMIS), by the Helmholtz Young Investigator Grant VH-NG-303,
and by the DFG Collaborative Research Grant SFB 676 
'Particles, Strings and the Early Universe'.

\appendix

\section{Determination of experimental $S_{95}$ values}
\label{S95-LEP}

\subsection{The $CL_s$ method}
In order to calculate the experimental $S_{95}$ values, the modified
frequentist method (or $CL_s$ method) is used. As a first step, using
the experimental data binned in $i$ bins with $d_i$ observed data
events, $s_i$ expected signal events for a given hypothesis of the
signal, and $b_i$ background events, a test statistics
$$Q=\prod_{i=1}^j \frac{e^{-(s_i+b_i)}(s_i+b_i)^{d_i}}
                      {d_i!}\left/\frac{e^{-(b_i)}(b_i)^{d_i}}{d_i!} \right.$$ 
is defined, which orders the expected outcomes of test
experiments according to their ``signal-likeness''. Then, confidence
levels for the given hypothesis are calculated. The confidence level
for the background hypothesis, ${\mathrm{CL}}_b$, is defined as the
probability to obtain values of a test statistics $Q$ no larger than
the observed value $Q_{\mathrm{obs}}$, given a large number of
hypothetical experiments with background processes only,
$$\mathrm{CL}_b=P(Q\le Q_{\mathrm{obs}}|{\mathrm{background}}).$$
Similarly, the confidence level for the signal+background hypothesis,
${\mathrm{CL}}_{s+b}$, is defined as the probability to obtain values
of $Q$ smaller than observed, given a large number of hypothetical
experiments with signal and background processes,
$${\mathrm{CL}}_{s+b}=P(Q\le
Q_{\mathrm{obs}}|{\mathrm{signal+background}}).$$
In principle,
${\mathrm{CL}}_{s+b}$ can be used to exclude the signal+background
hypothesis, given a model for Higgs boson production.  However, this
procedure may lead to the undesired possibility that a downward
fluctuation of the background would allow a hypotheses to be excluded
for which the experiment has no sensitivity due to the small expected
signal rate. This problem is avoided by introducing the ratio
$${\mathrm{CL}}_s={\mathrm{CL}}_{s+b}/{\mathrm{CL}}_b.$$
Since
${\mathrm{CL}}_b$ is a positive number less than one,
${\mathrm{CL}}_s$ will always be greater than ${\mathrm{CL}}_{s+b}$
and the limit obtained in this way will therefore be conservative.
This quantity is adopted for setting exclusion limits and a hypothesis
is considered to be excluded at the $95\,\%$ confidence level if the
corresponding value of ${\mathrm{CL}}_s$ is less than 0.05.  This
method is also called ``modified frequentist approach'', since not the
frequentistically determined confidence levels ${\mathrm{CL}}_b$ and
${\mathrm{CL}}_{s+b}$ but their ratio is used.

\subsection{Expected and observed confidence levels}
The expected confidence levels are obtained by replacing the observed
data configuration by a large number of simulated event configurations
for the two hypotheses background only or signal+background.  These
can be used to estimate the expected sensitivity of a search and to
compare the observed exclusion with the one expected with no signal
present. The expected $\mathrm{CL}_s$ is then the median of the
outcome of the large number of simulated experiments with background
only event configuration.

Once the experimentally observed and the expected values of
$\mathrm{CL}_s$ are derived as described above using the
hypothesis of a Higgs production in the channel under study with
SM strength (i.e. all BR and cross section scaling factors set
to 1), the observed and expected values of $S_{95}$ can be numerically
determined by varying the production rate of the channel by a scaling
factor $s$ until
$$\mathrm{CL}_s=0.05$$ is reached. The scaling factor $s$ is then
interpreted as observed or expected $S_{95}$, depending of whether the
observed or expected confidence level $\mathrm{CL}_s$ has been used.

The numerical calculation at LEP has been done with tools described in
\cite{Bock:2004xz} and \cite{Junk:1999kv}, while the Tevatron results
use different methods described in the individual references.

\medskip

\clearpage
\newpage
\pagebreak


\begin{thebibliography}{00}





\bibitem{higgs-mechanism}
	P.~W.~Higgs,
  	Phys.\ Lett.\  {\bf 12} (1964) 132;
  	Phys.\ Rev.\ Lett.\  {\bf 13} (1964) 508;
	Phys.\ Rev.\  {\bf 145} (1966) 1156;\\
	F.~Englert and R.~Brout,
	Phys.\ Rev.\ Lett.\  {\bf 13} (1964) 321;\\
	G.~S.~Guralnik, C.~R.~Hagen and T.~W.~B.~Kibble,
	Phys.\ Rev.\ Lett.\  {\bf 13} (1964) 585.

\bibitem{sm} S.L.~Glashow, 
             {\em Nucl.\ Phys.} {\bf B 22} (1961) 579; \\
             S. Weinberg, 
             {\em Phys. Rev. Lett.} {\bf 19} (1967) 19; \\ 
             A. Salam, in: {\em Proceedings of the 8th Nobel 
             Symposium}, Editor N. Svartholm, Stockholm, 1968.

\bibitem{mssm} H.~Nilles, 
               {\em Phys.\ Rept.} {\bf 110} (1984) 1; \\ 
               H.~Haber and G.~Kane, 
               {\em Phys.\ Rept.} {\bf 117} (1985) 75; \\  
               R.~Barbieri, 
               {\em Riv.\ Nuovo Cim.} {\bf 11} (1988) 1. 

\bibitem{thdm} S.~Weinberg,
	Phys.\ Rev.\ Lett.\  {\bf 37} (1976) 657;\\
	J.~Gunion, H.~Haber, G.~Kane and S.~Dawson,
        {\em The Higgs Hunter's Guide}
        (Perseus Publishing, Cambridge, MA, 1990),
        and references therein.

\bibitem{NMSSM-etc}
	P.~Fayet, 
        {\em Nucl. Phys.} {\bf B 90} (1975) 104;
	{\em Phys. Lett.} {\bf B 64} (1976) 159;
	{\em Phys. Lett.} {\bf B 69} (1977) 489;
	{\em Phys. Lett.} {\bf B 84} (1979) 416;\\
	H.P.~Nilles, M.~Srednicki and D.~Wyler,
	{\em Phys. Lett.} {\bf B 120} (1983) 346;\\
	J.M.~Frere, D.R.~Jones and S.~Raby,
	{\em Nucl. Phys.} {\bf B 222} (1983) 11;\\
	J.P.~Derendinger and C.A.~Savoy, 
        {\em Nucl. Phys.} {\bf B 237} (1984) 307;\\
	J.~Ellis, J.~Gunion, H.~Haber, L.~Roszkowski and F.~Zwirner,
	{\em Phys. Rev.} {\bf D 39}  (1989) 844;\\
	M.~Drees, 
        {\em Int. J. Mod. Phys.} {\bf A 4}  (1989) 3635;\\
	%
	U.~Ellwanger, M.~Rausch de~Traubenberg and C.A.~Savoy,
	{\em Phys. Lett.} {\bf B 315} (1993) 331 
        [arXiv:hep-ph/9307322];
	{\em Z. Phys.} {\bf C 67} (1995) 665 
        [arXiv:hep-ph/9502206];
	{\em Nucl. Phys.} {\bf B 492} (1997) 307 
        [arXiv:hep-ph/9611251];\\
	%
	T.~Elliott, S.F.~King and P.~White,
	{\em Phys.\ Lett.} {\bf B 351} (1995) 213 
        [arXiv:hep-ph/9406303];\\
	S.F.~King and P.~White,
	{\em Phys. Rev.} {\bf D 52} (1995) 4183 
        [arXiv:hep-ph/9505326];\\
	C.~Panagiotakopoulos and K.~Tamvakis,
	{\em Phys.\ Lett.} {\bf B 469} (1999) 145
	[arXiv:hep-ph/9908351];\\
	C.~Panagiotakopoulos and A.~Pilaftsis,
	{\em Phys.\ Rev.} {\bf D 63} (2001) 055003
	[arXiv:hep-ph/0008268];\\
	A.~Dedes, C.~Hugonie, S.~Moretti and K.~Tamvakis,
	{\em Phys.\ Rev.} {\bf D 63} (2001) 055009
	[arXiv:hep-ph/0009125].



\bibitem{lhm} N.~Arkani-Hamed, A.~Cohen and H.~Georgi,
              {\em Phys.\ Lett.} {\bf B 513} (2001) 232
              [arXiv:hep-ph/0105239];\\
              N.~Arkani-Hamed, A.~Cohen, T.~Gregoire and J.~Wacker,
              {\em JHEP} {\bf 0208} (2002) 020
              [arXiv:hep-ph/0202089].

\bibitem{edm} N.~Arkani-Hamed, S.~Dimopoulos and G.~Dvali,
              {\em Phys.\ Lett.} {\bf B 429} (1998) 263
              [arXiv:hep-ph/9803315];
              {\em Phys.\ Lett.} {\bf B 436} (1998) 257
              [arXiv:hep-ph/9804398];\\
              I.~Antoniadis,
              {\em Phys.\ Lett.} {\bf B 246} (1990) 377;\\
              J.~Lykken,
              {\em Phys.\ Rev.} {\bf D 54} (1996) 3693
              [arXiv:hep-th/9603133];\\
              I.~Antoniadis and M.~Quiros,
              {\em Phys.\ Lett.} {\bf B 392} (1997) 61
              [arXiv:hep-th/9609209];\\
              L.~Randall and R.~Sundrum,
              {\em Phys.\ Rev.\ Lett.} {\bf 83} (1999) 3370
              [arXiv:hep-ph/9905221];\\
	      D.~Cremades, L.~E.~Ibanez and F.~Marchesano,
  	      {\em Nucl.\ Phys.} {\bf B 643} (2002) 93
  	      [arXiv:hep-th/0205074];\\
	      C.~Kokorelis,
	      {\em Nucl.\ Phys.} {\bf B 677} (2004) 115
	      [arXiv:hep-th/0207234].




\bibitem{higgs-singlet-models}
	T.~Binoth and J.~J.~van der Bij,
	{\em Z.\ Phys.} {\bf C 75} (1997) 17
	[arXiv:hep-ph/9608245];\\
	D.~G.~Cerdeno, A.~Dedes and T.~E.~J.~Underwood,
	{\em JHEP} {\bf 0609} (2006) 067
	[arXiv:hep-ph/0607157];\\
	D.~O'Connell, M.~J.~Ramsey-Musolf and M.~B.~Wise,
	{\em Phys.\ Rev.} {\bf D 75} (2007) 037701
	[arXiv:hep-ph/0611014].

\bibitem{private-higgs}
	R.~A.~Porto and A.~Zee,
	arXiv:0712.0448 [hep-ph].

\bibitem{higgs-dep-Yuakawa} G.~F.~Giudice and O.~Lebedev,
  		{\em Phys.\ Lett.} {\bf B 665} (2008) 79
  		[arXiv:0804.1753 [hep-ph]].

\bibitem{LEP-SM-Higgs-analysis}
  R.~Barate et al.,
  {\em Phys.\ Lett.} {\bf B 565} (2003) 61
  [hep-ex/0306033].

\bibitem{LEP-MSSM-Higgs-analysis}
  S.~Schael et al.,
  {\em Eur.\ Phys.\ J.} {\bf C 47} (2006) 547
  [hep-ex/0602042].

\bibitem{LEPother} [LEP Higgs Working Group for Higgs boson searches]
  arXiv: hep-ex/0107031;
  arXiv: hep-ex/0107032;
  arXiv: hep-ex/0107033;
  arXiv: hep-ex/0107034;
  arXiv: hep-ex/0107035.


\bibitem{feynhiggs} S.~Heinemeyer, W.~Hollik and G.~Weiglein,
                    {\em Comput. Phys. Commun.} {\bf 124} (2000) 76,
                    [arXiv:hep-ph/9812320];
                    see: {\tt www.feynhiggs.de} .

\bibitem{mhiggslong} S.~Heinemeyer, W.~Hollik and G.~Weiglein,
                     {\em Eur. Phys. J.} {\bf C 9} (1999) 343
                     [arXiv:hep-ph/9812472].

\bibitem{mhiggsAEC} G.~Degrassi, S.~Heinemeyer, W.~Hollik,
                    P.~Slavich and G.~Weiglein, 
                    {\em Eur. Phys. J.} {\bf C 28} (2003) 133
                    [arXiv:hep-ph/0212020].

\bibitem{mhcMSSMlong} M.~Frank, T.~Hahn, S.~Heinemeyer, W.~Hollik,  
                      H.~Rzehak and G.~Weiglein,
                      {\em JHEP} {\bf 0702} (2007) 047
                      [arXiv:hep-ph/0611326].


\bibitem{cpsh} J.~S.~Lee, A.~Pilaftsis, M.~S.~Carena, S.~Y.~Choi, M.~Drees,
               J.~R.~Ellis and C.~E.~M.~Wagner, 
               {\em Comput.\ Phys.\ Commun.} {\bf 156} (2004) 283
               [arXiv:hep-ph/0307377];\\
               J.~S.~Lee, M.~Carena, J.~Ellis, A.~Pilaftsis and C.~E.~M.~Wagner,
               arXiv:0712.2360 [hep-ph].

\bibitem{:2005ema}
  The ALEPH Collaboration, The DELPHI Collaboration, The L3 Collaboration, 
  The OPAL Collaboration, The SLD Collaboration, The LEP Electroweak Working
  Group and The SLD Electroweak and Heavy Flavour Groups,
  {\em Phys.\ Rept.} {\bf 427} (2006) 257.
  [arXiv:hep-ex/0509008].

\bibitem{Alcaraz:2006mx}
  J.~Alcaraz et al.  [ALEPH Collaboration and DELPHI Collaboration and
                  L3 Collaboration and ],
  arXiv:hep-ex/0612034.

\bibitem{CDF-3b-analysis1} 
	The CDF Collaboration, CDF note 8954.

\bibitem{CDF-3b-analysis2}
	The CDF Collaboration, CDF note 9284.

\bibitem{D0-0805-3556}
	The D$\O$ Collaboration, Report FERMILAB-PUB-08-142-E, 
	arXiv:0805.3556,
	submitted to Phys. Rev. Lett.

\bibitem{D0-0805-2491}
	The D\O\ Collaboration, 
	{\em Phys.\ Rev.\ Lett.} {\bf 101} (2008) 071804,
	arXiv:0805.2491 [hep-ex].

\bibitem{CDF-9071}
	The CDF Collaboration, CDF note 9071.

\bibitem{hdecay}
  A.~Djouadi, J.~Kalinowski and M.~Spira,
  {\em Comput.\ Phys.\ Commun.} {\bf 108} (1998) 56
  [arXiv:hep-ph/9704448].

\bibitem{TEV4LHCWG-Higgs-CS}
	T.~Hahn, S.~Heinemeyer, F.~Maltoni, G.~Weiglein and S.~Willenbrock,
        arXiv:hep-ph/0607308;\\
  	U.~Aglietti et al.,
  	arXiv:hep-ph/0612172;\\
        See also: {\tt http://maltoni.web.cern.ch/maltoni/TeV4LHC/}.

\bibitem{HJET}
  O.~Brein and W.~Hollik,
  {\em Phys.\ Rev.} {\bf D 68} (2003) 095006
  [arXiv:hep-ph/0305321];
  {\em Phys.\ Rev.} {\bf D 76} (2007) 035002
  [arXiv:0705.2744 [hep-ph]].

\bibitem{VBFatNLO}
  T.~Figy, C.~Oleari and D.~Zeppenfeld,
  {\em Phys.\ Rev.} {\bf D 68} (2003) 073005
  [arXiv:hep-ph/0306109].


\bibitem{commun-Read}
Alexander L. Read, private communication.

\bibitem{commun-Bechtle}
LEP Higgs Working Group, unpublished.

\bibitem{CDF-0807-4493}
	The CDF Collaboration, 
        arXiv:0807.4493 [hep-ex].

\bibitem{CDF-9475}
	The CDF Collaboration, CDF note 9475.

\bibitem{D0-5570}
	The D\O\ Collaboration, D0 note 5570.

\bibitem{D0-5472}
	The D\O\ Collaboration, D\O\ note 5472.

\bibitem{CDF-9463}
        The CDF Collaboration, CDF note 9463.

\bibitem{D0-5485}
	The D\O\ Collaboration, D\O\ note 5485.




\bibitem{CDF-7307}
	The CDF Collaboration, CDF note 7307.



\bibitem{D0-5757}
        The D\O\ Collaboration, D\O\ note 5757.

\bibitem{CDF-0809-3930}
	T.~Aaltonen et al. [CDF Collaboration],
        arXiv:0809.3930 [hep-ex].


\bibitem{D0-0803-1514}
	The D\O\ Collaboration, 
	{\em Phys.\ Rev.\ Lett.} {\bf 101} (2008) 051801
  	[arXiv:0803.1514 [hep-ex]].

\bibitem{D0-5737}
        The D\O\ Collaboration, D\O\ note 5737.

\bibitem{D0-5726}
        The D\O\ Collaboration, D\O\ note 5726.

\bibitem{CDF-9483}
	The CDF Collaboration, CDF note 9483.

\bibitem{D0-5586}
	The D\O\ Collaboration, D\O\ note 5586.


\bibitem{CDF-9248}
        The CDF Collaboration, CDF note 9248.


\bibitem{CDF-D0-combined-1}
  TEVNPH Working Group Collaboration 
  for the CDF Collaboration and D0 Collaboration,
  Report FERMILAB-PUB-07-656-E,
  CDF note 8961, D\O\ note 5536,
  arXiv:0712.2383 [hep-ex].

\bibitem{CDF-D0-combined-2}
  TEVNPH Working Group Collaboration 
  for the CDF Collaboration and D0 Collaboration,
  Report FERMILAB-PUB-08-069-E,
  CDF note 9290, D\O\ note 5645,
  arXiv:0804.3423 [hep-ex].

\bibitem{CDF-D0-combined-3}
  TEVNPH Working Group Collaboration 
  for the CDF Collaboration and D0 Collaboration,
  Report FERMILAB-PUB-08-270-E,
  CDF note 9465, D\O\ note 5754,
  arXiv:0808.0534 [hep-ex].


\bibitem{SM-gluon-fusion-CS}
  D.~Graudenz, M.~Spira and P.~Zerwas,
  {\em Phys.\ Rev.\ Lett.} {\bf 70} (1993) 1372;\\
  M.~Spira, A.~Djouadi, D.~Graudenz and P.~Zerwas,
  {\em Nucl.\ Phys.} {\bf B 453} (1995) 17
  [arXiv:hep-ph/9504378];\\
  R.~Harlander and W.~Kilgore,
  {\em Phys.\ Rev.\ Lett.} {\bf 88} (2002) 201801
  [arXiv:hep-ph/0201206];\\
  V.~Ravindran, J.~Smith and W.~van Neerven,
  {\em Nucl.\ Phys.} {\bf B 665} (2003) 325
  [arXiv:hep-ph/0302135];\\
  C.~Anastasiou and K.~Melnikov,
  {\em Nucl.\ Phys.} {\bf B 646} (2002) 220
  [arXiv:hep-ph/0207004];\\
  S.~Moch and A.~Vogt,
  {\em Phys.\ Lett.} {\bf B 631} (2005) 48
  [arXiv:hep-ph/0508265];\\
  E.~Laenen and L.~Magnea,
  {\em Phys.\ Lett.} {\bf B 632} (2006) 270
  [arXiv:hep-ph/0508284].

\bibitem{SM-gluon-fusion-CS-Catani-etal}
  S.~Catani, D.~de Florian, M.~Grazzini and P.~Nason,
  {\em JHEP} {\bf 0307} (2003) 028
  [arXiv:hep-ph/0306211];\\

\bibitem{Harlander:2003ai}
  R.~V.~Harlander and W.~B.~Kilgore,
  {\em Phys.\ Rev.} {\bf D 68} (2003) 013001
  [arXiv:hep-ph/0304035].

\bibitem{Djouadi-anatomyI}
  A.~Djouadi,
  {\em Phys.\ Rept.} {\bf 457} (2008) 1
  [arXiv:hep-ph/0503172].

\bibitem{MSTW2008}
  A.~D.~Martin, W.~J.~Stirling, R.~S.~Thorne and G.~Watt,
  arXiv:0901.0002 [hep-ph].

\bibitem{langenegger-etal} 
U.~Langenegger, M.~Spira, A.~Starodumov and P.~Trueb,
  {\em JHEP} {\bf 0606} (2006) 035
  [arXiv:hep-ph/0604156].

\bibitem{SM-VH-CS}
  O.~Brein, A.~Djouadi and R.~Harlander,
  {\em Phys.\ Lett.} {\bf B 579} (2004) 149
  [arXiv:hep-ph/0307206];\\
  M.~L.~Ciccolini, S.~Dittmaier and M.~Kramer,
  {\em Phys.\ Rev.} {\bf D 68} (2003) 073003
  [arXiv:hep-ph/0306234];\\
  K.~A.~Assamagan et al.  [Higgs Working Group Collaboration],
  arXiv:hep-ph/0406152.


\bibitem{SM-VBF-CS}
  T.~Han, G.~Valencia and S.~Willenbrock,
  {\em Phys.\ Rev.\ Lett.} {\bf 69} (1992) 3274
  [arXiv:hep-ph/9206246];\\
  J.~M.~Campbell and R.~K.~Ellis,
  {\em Phys.\ Rev.} {\bf D 60} (1999) 113006
  [arXiv:hep-ph/9905386];\\
  T.~Figy, C.~Oleari and D.~Zeppenfeld,
  {\em Phys.\ Rev.} {\bf D 68} (2003) 073005
  [arXiv:hep-ph/0306109];\\
  E.~Berger and J.~Campbell,
  {\em Phys.\ Rev.} {\bf D 70} (2004) 073011
  [arXiv:hep-ph/0403194].

\bibitem{new-gluon-fusion-prediction}
  C.~Anastasiou, R.~Boughezal and F.~Petriello,
  arXiv:0811.3458 [hep-ph].

\bibitem{new-gluon-fusion-prediction-using}
U.~Aglietti, R.~Bonciani, G.~Degrassi and A.~Vicini,
  {\em Phys.\ Lett.}  {\bf B 595} (2004) 432
  [arXiv:hep-ph/0404071],
  arXiv:hep-ph/0610033;\\
S.~Actis, G.~Passarino, C.~Sturm and S.~Uccirati,
  {\em Phys.\ Lett.} {\bf B 670} (2008) 12
  [arXiv:0809.1301 [hep-ph]];\\
C.~Anastasiou, S.~Beerli, S.~Bucherer, A.~Daleo and Z.~Kunszt,
  {\em JHEP} {\bf 0701} (2007) 082
  [arXiv:hep-ph/0611236].

\bibitem{MRST2006}
  A.~D.~Martin, W.~J.~Stirling, R.~S.~Thorne and G.~Watt,
  {\em Phys.\ Lett.} {\bf 652} (2007) 292
  [arXiv:0706.0459 [hep-ph]].

\bibitem{extra-gen-review}
  P.~H.~Frampton, P.~Q.~Hung and M.~Sher,
  {\em Phys.\ Rept.} {\bf 330} (2000) 263
  [arXiv:hep-ph/9903387].

\bibitem{four-gen-and-Higgs}
  G.~D.~Kribs, T.~Plehn, M.~Spannowsky and T.~M.~P.~Tait,
  {\em Phys.\ Rev.} {\bf D 76} (2007) 075016
  [arXiv:0706.3718 [hep-ph]].

\bibitem{MSSMHiggsRev} S.~Heinemeyer, 
  {\em Int.\ J.\ Mod.\ Phys.} {\bf A 21} 2659 (2006)
  [arXiv:hep-ph/0407244];\\
  A.~Djouadi, 
  {\em Phys.\ Rept.} {\bf 459} (2008) 1.
  arXiv:hep-ph/0503173.

\bibitem{Hahn:2006np}
  T.~Hahn, S.~Heinemeyer, W.~Hollik, H.~Rzehak, G.~Weiglein and K.~Williams,
  {\em Pramana} {\bf 69} (2007) 861
  [arXiv:hep-ph/0611373].

\bibitem{benchmark2} M.~Carena, S.~Heinemeyer, C.~Wagner and G.~Weiglein, 
                     {\em Eur. Phys. J.} {\bf C 26} (2003) 601
                     [arXiv:hep-ph/0202167].

\bibitem{mt1724} Tevatron Electroweak Working Group and CDF Collaboration
                 and D0 Collaboration,
                 arXiv:0808.1089 [hep-ex].

\bibitem{tbexcl} S.~Heinemeyer, W.~Hollik and G.~Weiglein, 
                 {\em JHEP} {\bf 0006} (2000) 009
                 [arXiv:hep-ph/9909540].

\bibitem{Carena:2000ks}
  M.~S.~Carena, J.~R.~Ellis, A.~Pilaftsis and C.~E.~M.~Wagner,
  {\em Phys.\ Lett.} {\bf B 495} (2000) 155
  [arXiv:hep-ph/0009212].

\bibitem{Carena:2000yi}
  M.~S.~Carena, J.~R.~Ellis, A.~Pilaftsis and C.~E.~M.~Wagner,
  {\em Nucl.\ Phys.} {\bf B 586} (2000) 92
  [arXiv:hep-ph/0003180].

\bibitem{Williams:2007dc}
  K.~E.~Williams and G.~Weiglein,
  {\em Phys.\ Lett.} {\bf B 660} (2008) 217
  [arXiv:0710.5320 [hep-ph]].


\bibitem{Bock:2004xz}
  P.~Bock,
  {\em JHEP} {\bf 0701} (2007) 080
  [arXiv:hep-ex/0405072].


\bibitem{Junk:1999kv}
  T.~Junk,
  {\em Nucl.\ Instrum.\ Meth.} {\bf A 434} (1999) 435
  [arXiv:hep-ex/9902006].

\end{thebibliography}
\end{document}